\documentclass[8pt,preprint2]{aastex}

\shorttitle{Red Sequence Luminosity Function}
\shortauthors{Crawford et al.}

\begin{document}

\title {The Red Sequence Luminosity Function in Massive Intermediate
  Redshift Galaxy Clusters }

\author{S. M. Crawford,\altaffilmark{1} M. A. Bershady,\altaffilmark{2}
and J. G. Hoessel$^2$}

\altaffiltext{1}{South African Astronomical Observatory, Observatory, 7935 Cape Town, South Africa; crawford@saao.ac.za}
\altaffiltext{2}{Washburn Observatories, U. Wisconsin - Madison, 475 N. Charter St., Madison, WI 53706; mab@astro.wisc.edu} 

\begin{abstract}

We measure the rest-frame B-band luminosity function of red-sequence
galaxies (RSLF) of five intermediate-redshift ($0.5 < z < 0.9$),
high-mass ($\sigma > 950$ km s$^{-1}$) clusters. Cluster galaxies are
identified through photometric redshifts based on imaging in seven
bands (five broad, and two narrow) using the WIYN 3.5m telescope.  The
luminosity functions are well-fit down to $M_B^*+3$ for all of the
clusters out to a radius of $R_{200}$.  For comparison, the luminosity
functions for a sample of 59 low redshift clusters selected from the
SDSS are measured as well.  There is a brightening trend ($M_B^*$
increases by 0.7 mags by z=0.75) with redshift comparable to what is
seen in the field for similarly defined galaxies, although there is a
hint that the cluster red-sequence brightening is more rapid in the
past ($z>0.5$), and relatively shallow at more recent times.  Contrary
to other claims, we find little evidence for evolution of the faint
end slope.  Previous indications of evolution may be due to
limitations in measurement technique, bias in the sample selection,
and cluster to cluster variation.  As seen in both the low and high
redshift sample, a significant amount of variation in luminosity
functions parameters $\alpha$ and $M^*$ exists between individual
clusters.

\end{abstract}


\section{Introduction}

In the nearby Universe, galaxy clusters are dominated by early type
galaxies (Dressler 1980) along the so-called ``red sequence,'' the
name for which derives from the tight color-magnitude relation
observed for these galaxies (Visvanathan \& Sandage 1977).  The most
luminous cluster galaxies, located at the tip of the red sequence,
appear to have relatively little star formation since $z \sim 3$
(Bower et al. 1992; Ellis et al. 1997; Stanford et al.  1998; Kelson
et al. 2001).  Observations of high redshift ($z>1$) clusters are
consistent with this inference, indicating at least the bright end of
the color-magnitude relationship has only passively evolved since
$z\sim1.5$ (Stanford et al. 1997, Mullis et al. 2005, Stanford et
al. 2005). However, evidence for recent ($\leq 5$ Gyr) bursts of star
formation in today's lower-mass cluster galaxies (Poggianti et
al. 2001, Conselice et al. 2003) indicates not all cluster galaxies
have the same homogeneous star formation histories. While the luminous
red sequence in clusters may have long been in place, the extension of
the red sequence to lower luminosity may be a relatively recent
phenomenon.

Recent, deep imaging programs have led to the claim of a deficit of
red sequence at low luminosities in high redshift clusters (Nakata et
al. 2001, de Lucia et al. 2004, Goto et al. 2005, Tanaka et
al. 2005). This would suggest many of the galaxies seen along today's
red sequence may be late additions, perhaps culled from a quenched and
faded blue population. This fits neatly with a notion that the red
sequence is not monolithic, as evinced by the seminal studies of
the Virgo cluster luminosity function (Sandage, Binggeli \& Tammann
1985). However, the deficit may not be ubiquitous (e.g., Andreon
2006).  Differences in the analysis methods, clusters samples, and the
inherent limitations in different data sets (i.e., the depth of
available data) have left the observational evidence for a deficit
inconclusive.

Differences between studies are also complicated by the possible
existence of evolutionary phenomenon that is a function of cluster
mass.  Substantial differences exist between the cluster and field red
sequences (de Propris et al. 2003, Croton et al. 2005), and there is
evidence for differences in the luminosity function with cluster mass,
both locally (Hansen et al. 2005, Hilton et al. 2005) and at
intermediate redshift (Koyama et al. 2007, Gilbank et al. 2007).
However, evidence to the contrary has also appeared in the literature
(Barkhouse et al. 2007, Andreon 2007). Given these disparate claims,
it is difficult to assemble a clear picture of the red-sequence
evolution -- a challenge again exacerbated by differences in cluster
samples and measurement techniques between studies.

A common theme repeated in many of the studies is the large cluster to
cluster variation seen in any of the parameterizations of the cluster
populations, particularly the luminosity function.  Estimates of the
dwarf-to-giant ratio (DGR, de Lucia et al. 2006) and Schechter-function
(Schechter 1976) fits to the luminosity function (Barkhouse et
al. 2007, Andreon 2007) both show significant cluster-to-cluster
variation. Unique features also have long been identified in the shape
of luminosity functions of local clusters (Biviano et al. 1995, Yagi
et al. 2002), indicative of a multi-component population even among
the red-galaxy population.  Traditionally, deep studies of the
so-called red-sequence luminosity function (RSLF) in local clusters
have fit two functions (either double Schechter functions or a Gaussian
and Schechter function) to the distribution of galaxies (Sandage et al.
1985, Jerjen \& Tammann 1997, Popesso et al. 2005). The bright end of
the luminosity function is well described by a functional form with a
sharp turnover at lower luminosities, while the fainter
luminosity-function component is generally found to be rapidly rising
at lower luminosities. Variations in the relative amplitude of these
components may well drive overall variations in observed cluster
luminosity functions. Studies of intermediate redshift clusters are
dominated by this brighter component, due to depth limitations. While
such studies rarely reach deep enough to well-characterize the
rapidly-rising faint end, its modulation may significantly impact the
observed count-parametrization.

In this paper, we explore the red-sequence population in five
high-density regions containing massive, bona fide, and well-studied
clusters between $0.5<z < 0.9$. Our study is based on deep, multi-band
imaging data from the WIYN 3.5m telescope and extant spectroscopic
data. The depth of our WIYN data is comparable to other cluster
studies at similar redshifts, and allows an independent assessment of
the cluster RSLF significantly below the knee in the bright end of the
luminosity function. The extent to which this, or any other extant
study at intermediate redshift, can address the faint-end of the
cluster RSLF is a point we address. For comparison, we include
measurements of the luminosity function for a large sample of low
redshift clusters from the Sloan Digital Sky Survey measured in the
same manner as our intermediate redshift sample.

In \S 2 we summarize the observation data and highlight our
measurement techniques.  In \S \ref{sec:lf}, we present the luminosity
functions measured for each of the clusters as well as the tests to
confirm and assess the reliability of the measured luminosity
functions. In \S \ref{sec:lowz}, the low-redshift cluster sample is
described. Finally, we discuss the observed evolution in our
luminosity functions (\S 4), compare to results in the literature, and
discuss the implications of our results in \S 5.  Throughout this
work, we adopt H$_{0}=70$ km s$^{-1}$ Mpc$^{-1}$, $\Omega_{Matter} =
0.3$, and $\Omega_{\Lambda} = 0.7$; all absolute magnitudes are in the
rest-frame (Johnson) B-band in the Vega system.

\section{Observations and Analysis \label{sec:obs}}

Observations were obtained between 1999 October and 2004 June, with
the WIYN\footnote{The WIYN Observatory is a joint facility of the
University of Wisconsin-Madison, Indiana University, Yale University,
and the National Optical Astronomy Observatories.} 3.5m telescope's
Mini-Mosaic Camera ($0.14^{\prime\prime}$ per pixel and $9.6^\prime
\times 9.6^\prime$ field of view, $0.85^{\prime\prime}$ FWHM median
seeing) as part of the survey described by Crawford et al. (2006) and
Crawford (2006). Cluster images were taken in the Harris UBRI, Gunn z
(Schneider, Gunn, \& Hoessel 1983), and two narrow band filters. The
narrow band filters ($\lambda/\Delta\lambda\sim70$) were specifically
designed to detect [OII]$\lambda 3727$ and adjacent continuum at each
cluster redshift.  Table \ref{tab:clust} lists the five
clusters observed, and their salient attributes.

Data analysis proceeded on deep mosaic images created from reduced
Mini-Mo frames, flat to $1\%$ of their initial sky values.  The deep
mosaics were photometrically calibrated through a variety of
independent methods yielding uncertainties below $2\%$.  Object
detection was performed on the R-band (typically deepest) images using
SExtractor (Bertin \& Arnouts 1996) with the criterion that objects
contained $>20$ contiguous pixels above $3\sigma$ of the sky
noise. This corresponds to a minimum signal-to-noise (S/N) of 13.4 in
an equivalent circular aperture of 0.35 arcsec radius (just slightly
smaller than the typical seeing disk). The S/N at the 50\% detection
limit (see below) will be higher since most objects intrinsically at
the specified limiting size and surface-brightness will have some
fraction of the pixellated signal noise-aberrated below the detection
threshold. As the S/N depends on the aperture size and tends to peak
around the the half-light radius, it is essential to define the S/N
used in any analysis to avoid confusion.

Two types of magnitudes were used for all calculations.
Seeing-matched\footnote{All images were degraded to worst-case seeing
for each cluster for the purpose of this photometric measurement.}
aperture magnitudes with a radius corresponding to 7.5 kpc at each
cluster redshift were used for all colors and photometric redshifts.
This is equivalent to aperture diameters between 1.93 and 2.45 arcsec,
or roughly $2.5\times$ stellar FWHM on average. This aperture defines
the S/N value for the R-band that we use in subsequent analysis,
particularly in the next two sub-sections concerning detection
completeness and photometric-redshift errors. Tailored magnitudes
(Crawford et al. 2006) based on the curve of growth, concentration
index, and Petrosian ratio are used for total apparent and absolute
magnitudes.  Absolute magnitudes and colors were calculated for all
objects using photometric (see below) or spectroscopic
redshifts. K-correction calculations adopt method 4 of Bershady
(1995).

\subsection{Detection Completeness}

\begin{figure}[hbt]
\epsscale{0.9}
\plotone{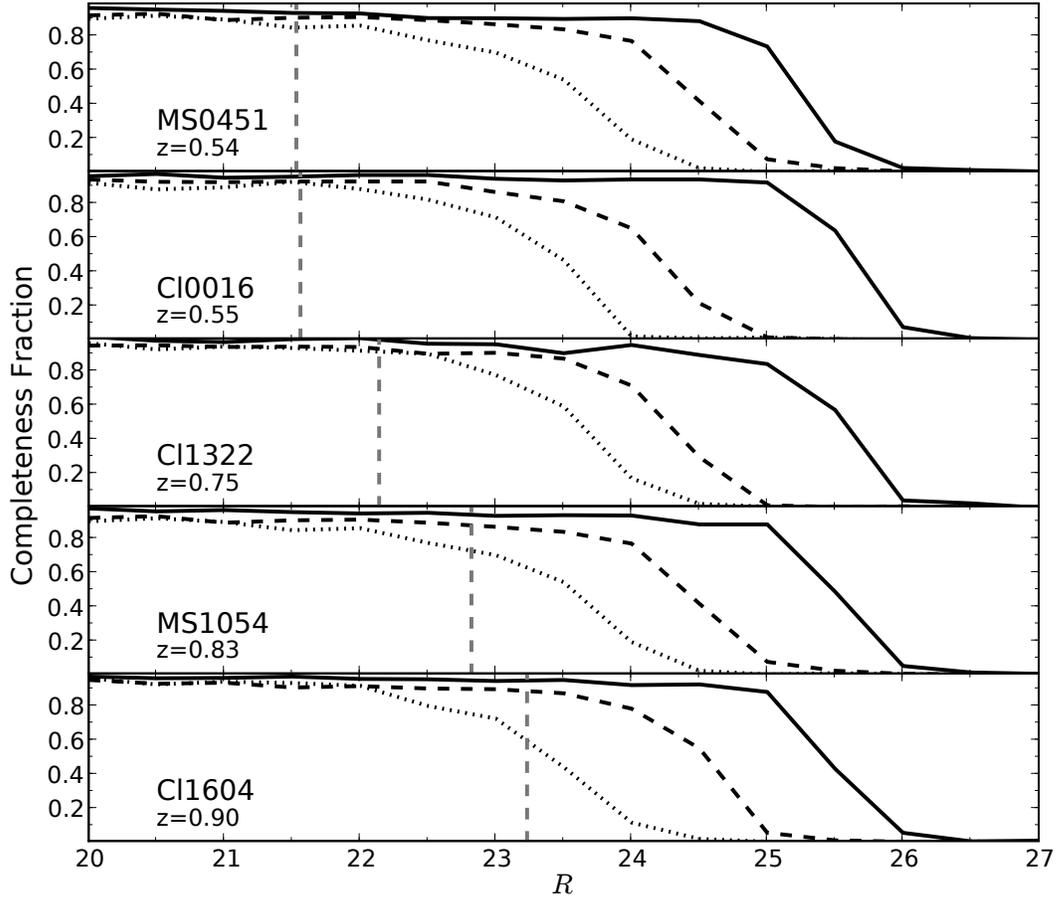}
\caption{R-band completeness profiles as function of object size for
  the six fields observed for this program. Curves (dotted, dashed,
  solid) are measured for real objects selected from the data with
  apparent half-light sizes of approximately 1.5',1', and 0.5',
  respectively. For reference, the approximate location of M* for the
  evolving field RSLF (Willmer et al. 2006) is shown as a vertical
  line.}
\label{fig:comp}
\end{figure}

We determine the completeness limit of our data through Monte Carlo
simulations similar to what is described in Bershady, Lowenthal \& Koo
(1995). In each detection image, we extract a sample of bright,
representative objects of different sizes.  Objects are dimmed, and
then re-inserted into the image. For each half-magnitude interval, we
insert 50 objects randomly across the usable field of view, repeat the
detection procedure, and repeat the process 10 times, and for a
minimum of three different object sizes. For small, unresolved
objects, the $50\%$ completeness magnitude is typically around
$R\sim25.5$ magnitudes. For objects with lower surface brightness, our
completeness limits can be significantly ($\sim1$ mag) more
shallow. The R-band completeness curves are presented in Figure
\ref{fig:comp}. For guidance as to the depth of these data, we mark in
the Figure the location of M* at the appropriate cluster redshift
based on the evolving RSLF estimated for the field (Willmer et al.
2006). When correcting for incompleteness, we calculate the correction
relative to the measured size of each object. S/N in seeing-matched
apertures at the 50\% and 90\% detection completeness are roughly 20
and 25, respectively, which is well above the limiting S/N for
detection.

For luminous red galaxies, the completeness limit is relatively
shallow due to the diffuse profile shape and large size of such
galaxies. Although a significant portion of the light is concentrated
into the central regions, most of the light is in the low surface
brightness wings of the galaxy. However, these systems are not at the
detection-limit of our survey at the cluster redshifts. Instead,
intrinsically faint (low luminosity), red galaxies are sampled near
our detection limits, for which the above detrimental properties are
not in play.  Three natural phenomenon are working in our favor in
terms of detecting these lower-luminosity galaxies. First, the light
profiles of lower-luminosity red-sequence galaxies are closer to an
exponential distribution than an $r^{1/4}$ (Graham \& Guzm{\'a}n
2003), thereby enhancing detectability.  Second, according to the
luminosity-size distribution of red galaxies (Shen et al. 2003,
McIntosh et al. 2005), we expect the faint, red galaxies to be
unresolved in our data set, and hence our deepest detection limits are
relevant. We return to, and illustrate these two points later in \S 3.
Finally, lower-luminosity galaxies are expected to be slightly bluer
than their luminous red-sequence counterparts due to a slope in the
color-magnitude relationship. Hence, the effect of the k-correction
for these galaxies at the R-band detection limit is less severe.  On
this basis, we compute the depth of our survey relative to the $50\%$
completeness limit for unresolved objects in that field. These depths
are listed in Table \ref{tab:clust}, and relative to the field RSLF,
they range between 2.6 and 3.6 mag below $M^*$.
 
\subsection{Photometric Redshift Method, Accuracy and Precision}

\begin{figure}[hbt]
\epsscale{0.9}
\plotone{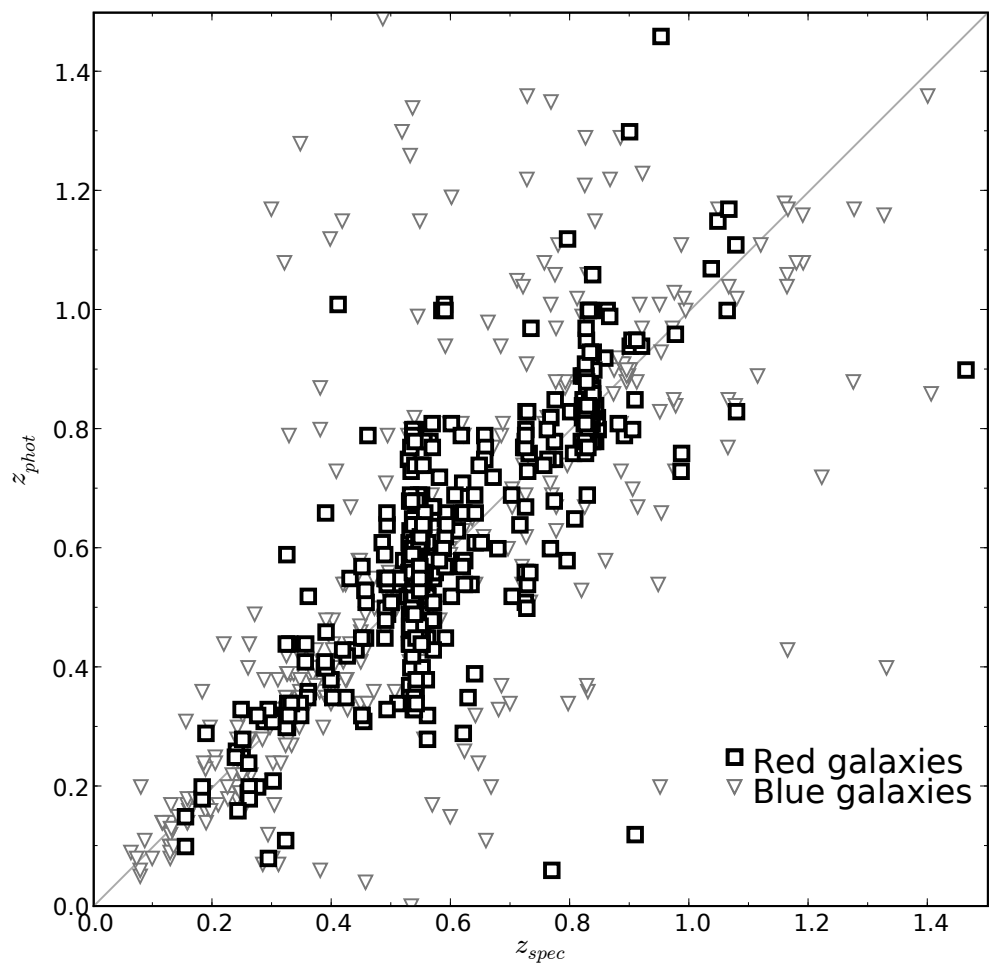}
\caption{Photometric redshift vs. spectroscopic redshift for the full
  sample of galaxies with spectroscopic redshifts in our survey
  fields.  Blue and red galaxies are defined in terms of rest-frame
  properties as specified in text.}
\label{fig:redz}
\end{figure}

\begin{figure}[hbt]
\epsscale{0.9}
\plotone{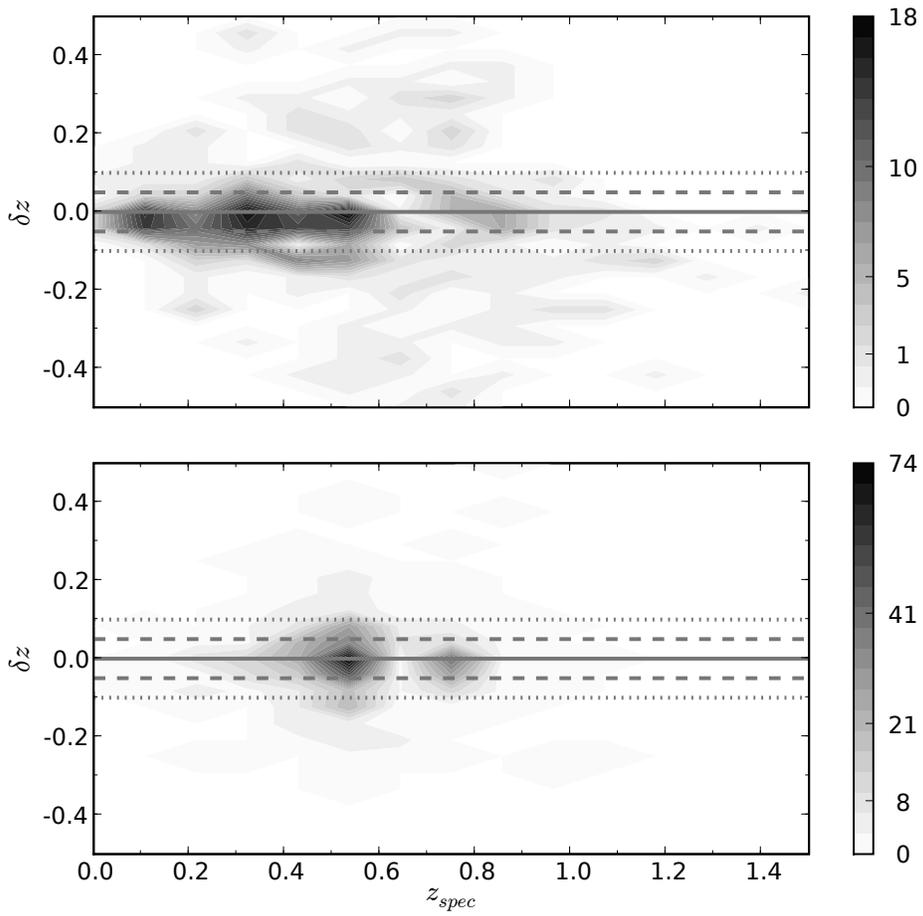}
\caption{Histogram of the difference between spectroscopic and
  photometric redshift as a function of z for the sample in Figure
  1. The top plot is for all blue galaxies and the bottom plot is for
  red galaxies, as defined in text.}
\label{fig:redzdiff}
\end{figure}

\begin{figure}[hbt]
\epsscale{0.9}
\plotone{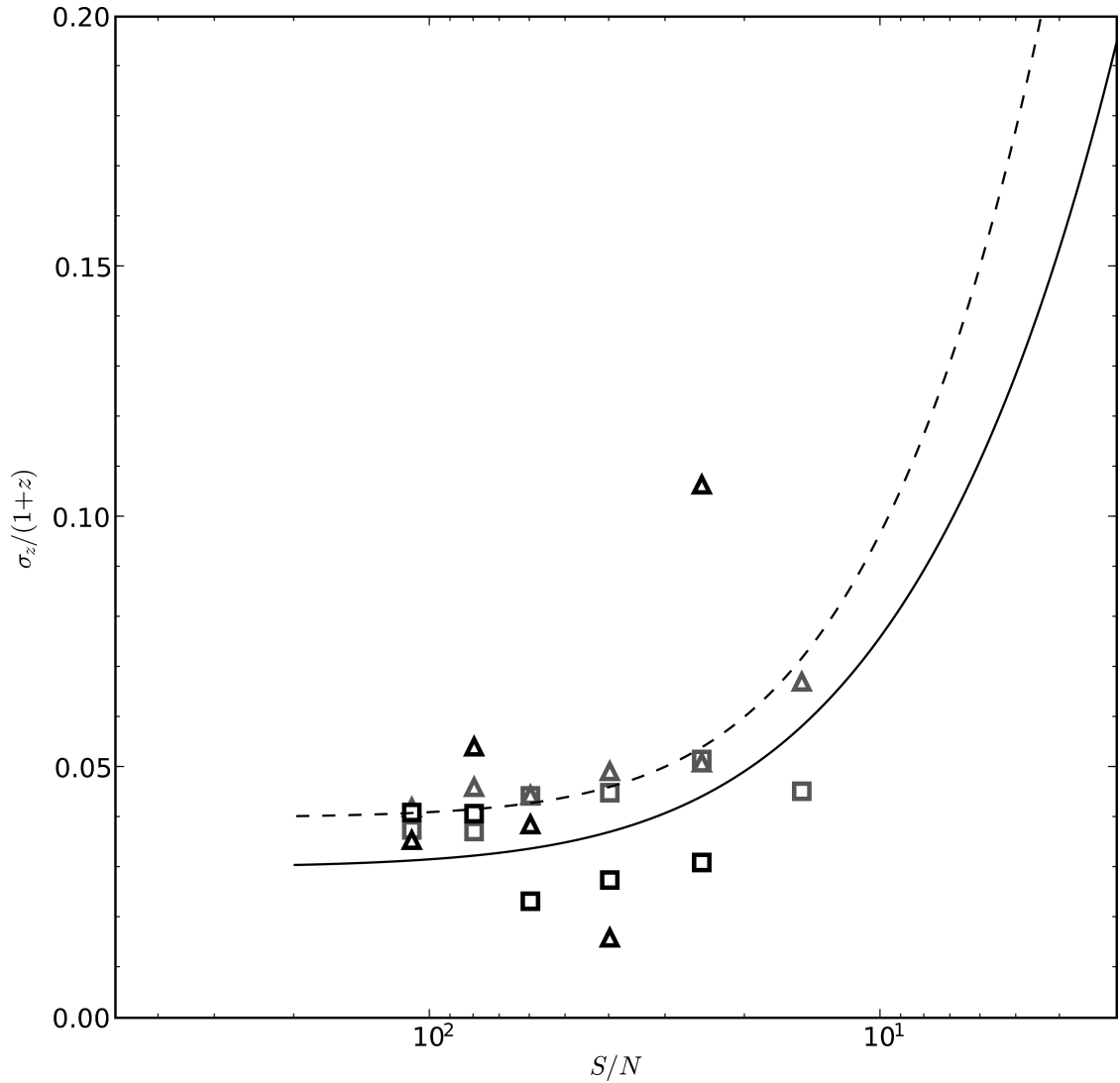}
\caption{Photometric redshift error as a function of S/N for
  measurements in this study. S/N is defined within the seeing-matched
  apertures described in \S 2. For the sample in Figure 1, the squares
  represent red galaxies; triangles represent blue galaxies. Light
  gray objects are similar measurements made from the SDSS sample
  using our photometric redshift technique. Photometric redshift error
  begins to increase dramatically below S/N=10, but sources at or
  above the 50\% detection limit have S/N $>$ 20.}
\label{fig:snz}
\end{figure}

We have derived photometric redshifts for all objects with the highest
precision possible.  The photometric redshifts were determined through
a hybrid of the template and training-set methods similar to Csabai et
al. (2003).  A template grid of model spectral energy distributions
(Bruzual \& Charlot 2003), and their corresponding broad- and
narrow-band fluxes, was created. The spectra cover a range of star
formation rates, ages, and redshifts.  Using objects with
spectroscopic redshifts and our observed multi-band fluxes, the grid
was constrained (more specifically, morphed or adjusted in flux space)
to match the measured data. Over 500 objects with existing
spectroscopic redshifts in our combined survey fields (Dressler et
al. 1997, Ellingson et al. 1998, Postman et al. 1998, van Dokkum et
al. 1999, Moran et al. 2007, Tran et al. 2007) trained our set of
templates in this way. The method was tested with the following
procedure. A single spectroscopic object was removed from the catalog.
The grid was trained using the remaining objects and then applied to
the removed objects.  The procedure was repeated so that an
independent measure of the photometric redshift (and its error) was
made for every spectroscopic object. In Figure
\ref{fig:redz} and \ref{fig:redzdiff}, we compare the spectroscopic
and photometric redshifts measured via this procedure. High quality
photometric redshifts can only be obtained with large values of signal
to noise as shown in Figure \ref{fig:snz}.  Photometric redshifts are
calculated with a precision $\sigma_z = 0.03$ and an accuracy of
$\delta_z < 0.01$ for red objects with $S/N > 30$ in the R-band. Blue
objects (defined below in \S2.3) performed slightly worse with
$\sigma_z = 0.05 $ and $\delta_z=0.01$ with $S/N > 30$.  Both types of
objects have much larger scatter for galaxies with $S/N < 30$.  At the
detection threshold, the typical scatter for red objects is expected
to be approximately $\sigma_z=0.05$.

\subsection{Impact on Luminosity Function Calculations}

Extensive simulations of the effect of photometric errors on the
distribution of measured {\it photometric-redshift} errors, as well as
the subsequent effect on the cluster-galaxy luminosity function were
performed. The latter comes about because we define cluster membership
based on photometric redshift (\S 3) when spectroscopic redshifts are
not available. Photometric redshift errors at high signal to noise
($S/N>10$) are small, and have a Gaussian distribution. At low signal
to noise, however, the error distribution is a combination of a
Gaussian core around the fiducial redshift plus a catastrophic error
component due to near-degeneracies in multi-color space. Combined with
our definition of cluster membership, the photometric-redshift errors
cause a significant portion ($>10\%$) of the cluster population above
our detection limit to be identified as field galaxies beyond our
redshift selection window. Because of large cluster over-densities,
contamination from the field into the cluster sample is
minimal. However, the opposite is true when we construct the
field-population estimate for the RSLF, as we show below in \S
2.3. Consequently, the nature of the correction is to account for a
net loss, or incompleteness in the bona-fide number of cluster
galaxies, and vice-versa for the field population. These corrections
are independent of our detection completeness except in so far as both
are correlated with apparent magnitude (or S/N). Statistical
corrections, based on our simulations, are applied to our luminosity
functions as a function of apparent magnitude. At the completeness
limit in the images, the average correction is $20\%$ with a maximum
of $40\%$ for CL0016.  For the field-galaxy luminosity function that
we construct below as a quality check, we push these corrections
farther to illustrate the quality of these corrections. In \S 3.2.1 we
discuss tests of the above issues of completeness and contamination in
the specific context of our cluster selection function.

\subsection{Quality Check:  The Field RSLF at Intermediate Redshifts}

\begin{figure}[hbt]
\epsscale{0.8}
\plotone{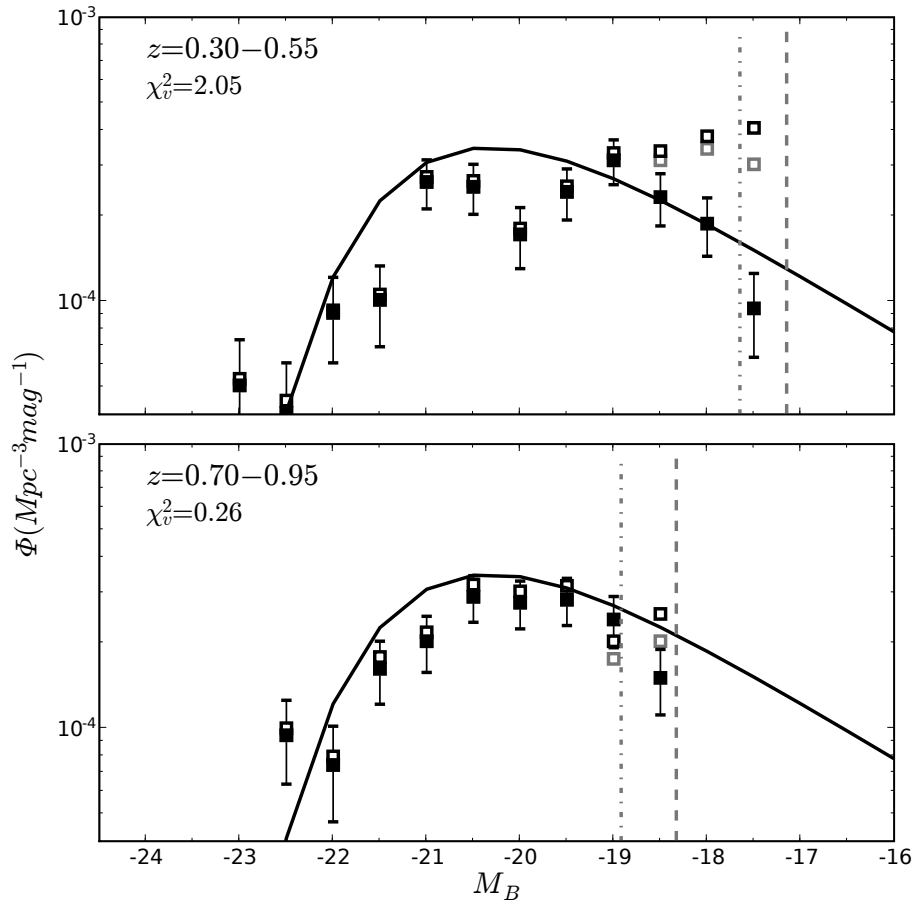}
\caption{Field RSLF at low (top panel; $0.3 < z < 0.55$) and high
  (bottom panel; $0.7 < z < 0.95$) redshift in our survey fields.
  Dashed and dotted vertical lines represent the estimated 50\% and
  90\% detection-completeness limits for our data in each redshift
  bin.  Light, open symbols represent raw (uncorrected) measurements
  in our images for redshift regimes free from over-densities (as
  defined in text). Dark, open symbols are corrected for detection
  completeness. Filled symbols are corrected for both detection
  completeness and cluster contamination due to photometric redshift
  efforts, as described in text. Unlike cluster luminosity-function
  derivations, here we extend the data beyond where contamination
  corrections are greater than $10\%$ for illustration purposes on the
  quality of the correction, which is based on simulations. For
  comparison, the field RSLF measured from the DEEP2 data set (Willmer
  et al. 2006) is plotted as a solid line. A reduced-$\chi2$ statistic
  between our data and the DEEP2 Schecter-function (no degrees of
  freedom) is indicated in each panel.}
\label{fig:fieldlf}
\end{figure}

To demonstrate our methodology for selecting and computing the
statistical distribution of red cluster galaxies, we first construct
the {\it field} RSLF luminosity function, derived from our data, in
two redshift bins that should be uncontaminated by rich,
over-densities, i.e., foreground and background regions in the cluster
fields (Figure \ref{fig:fieldlf}). The low redshift bin ($z=0.3-0.55$)
contains the imaging from fields MS1054, Cl1322, and Cl1604, whereas
the high redshift bin ($z=0.7-0.95$) is calculated based on data from
Cl0016 and MS0451.  In addition, one empty ``field'' image centered at
13:24:50.1h +30:11:18.5$^\circ$, which contains no evidence of any
massive over-density, was included in both bins. Red galaxies were
identified following the same prescription as Willmer et al. (2006):
\begin{equation} 
  U-B < -0.032*(M_B+21.52)+0.204.
\end{equation} 
This definition,
used throughout the paper, is based on a fit to the color-magnitude
relation plus a -0.25 mag shift in zeropoint.

Corrections for detection incompleteness and contamination due to
photometric redshift errors (as described in the previous section)
have been applied. The data are only presented to a magnitude limit
where the completeness is $>50\%$, but in some cases the contamination
corrections are larger.  At low S/N (faint magnitudes), cluster
sources with large photometric-redshift errors can contribute
substantially to the field counts.  Akin to the Eddington effect
(Eddington 1913) where fainter, more plentiful objects have a more
significant influence on brighter bins with less objects, the majority
of field counts at faint magnitudes can be due to contamination from
cluster over-densities.  For this reason, and to keep contamination
corrections manageable, we calculate the luminosity function after
excluding redshift bins where, for a given luminosity, the
contribution from the cluster is expected to be greater than $5\%$.
For example, at $R=25$ ($M_B=-18.6$) in the low-z sample, the error on
the photometric redshift is expected to scatter up to $10\%$ of the
galaxies from MS1054 into the redshift bin of $\Delta z = 0.5-0.6$.
Hence, in this magnitude bin, we would exclude galaxies with
photometric redshifts between $z=0.5-0.6$ in the MS1054 field from the
low-z luminosity function while still including galaxies with lower
redshifts.  An appropriate volume would be calculated for this
magnitude bin according the corresponding redshift window.
 
In Figure \ref{fig:fieldlf}, we compare our raw and corrected field
RSLF to DEEP2's Schechter-function parameterization of the same
(Willmer et al. 2006). A reduced-$\chi^2$ statistic of this
parameterization with respect to our data-set indicates a reasonable
match between results, despite our smaller survey volume and use of
photometric redshifts.  We reach a similar conclusion comparing our
field blue-galaxy luminosity function with corresponding results
Willmer et al. (2006). This is interesting because in the case of the
blue-galaxy luminosity, the field sample we construct has smaller
contamination errors.  On balance, these results indicate we are able
to control our corrections for incompleteness and photometric errors
reliably down to and below the level we will use in our cluster
luminosity function analysis.

\section{Cluster Red-Sequence Luminosity Functions \label{sec:lf}}

\subsection{WIYN Intermediate Redshift Sample}

\begin{figure}[hbt]
\epsscale{0.9}
\plotone{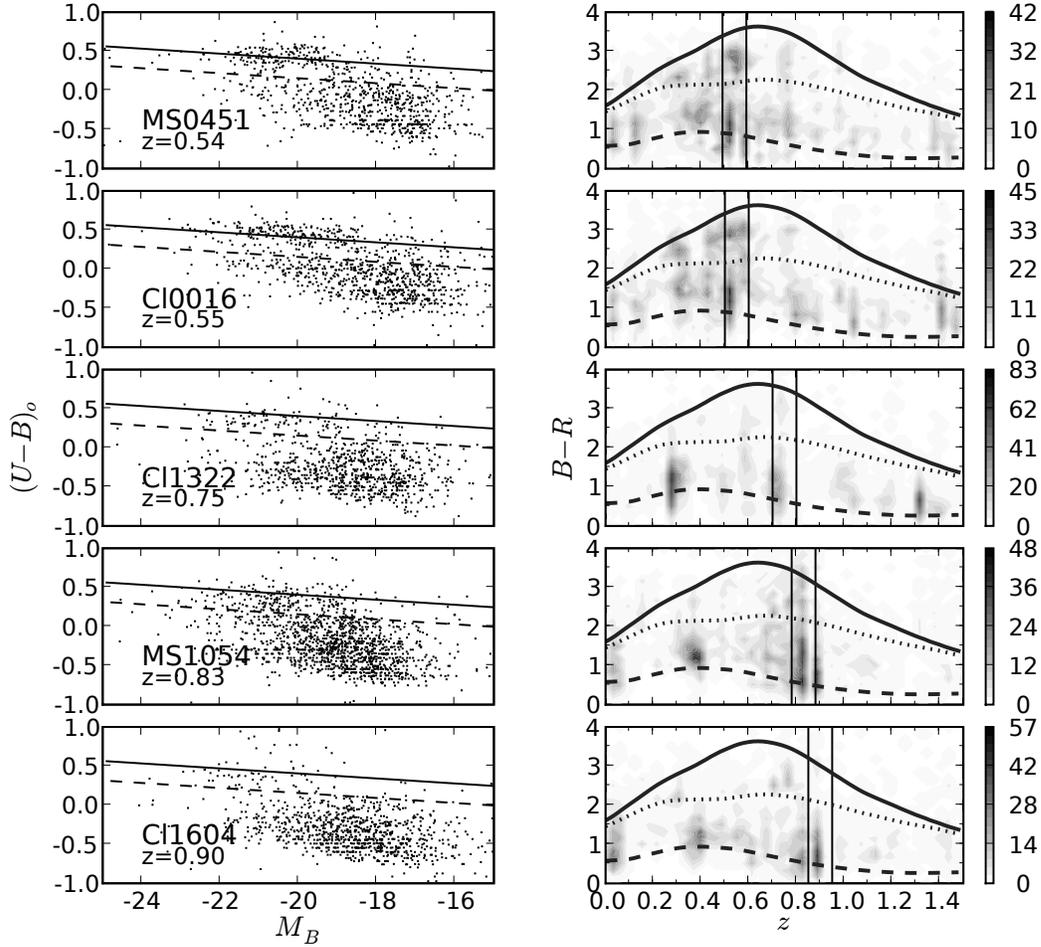}
\caption{(left) Rest-frame color-magnitude diagrams for each of the
  clusters. Only galaxies with photometric redshifts within $\pm 0.05$
  of the cluster redshift and within $R_{200}$ are plotted.  The
  redshift selections are illustrated in the right panels. Solid lines
  are the color-magnitude relationship from Willmer et al. 2006;
  dotted lines are the demarcation between red and blue galaxies
  (Equation 1 in text). (right) Apparent $B-R$ color vs redshift
  distribution for each cluster field. The solid vertical lines
  represent the redshift selection limits for each cluster, but all
  sources identified as galaxies in the field are plotted. The
  expected colors as a function of redshift for an non-evolving
  ellipitical galaxy (solid), Sb galaxy (dotted), and NGC4449 (dashed)
  are overplotted.}
\label{fig:cmzhist}
\end{figure}

\begin{figure}[hbt]
\epsscale{0.9}
\plotone{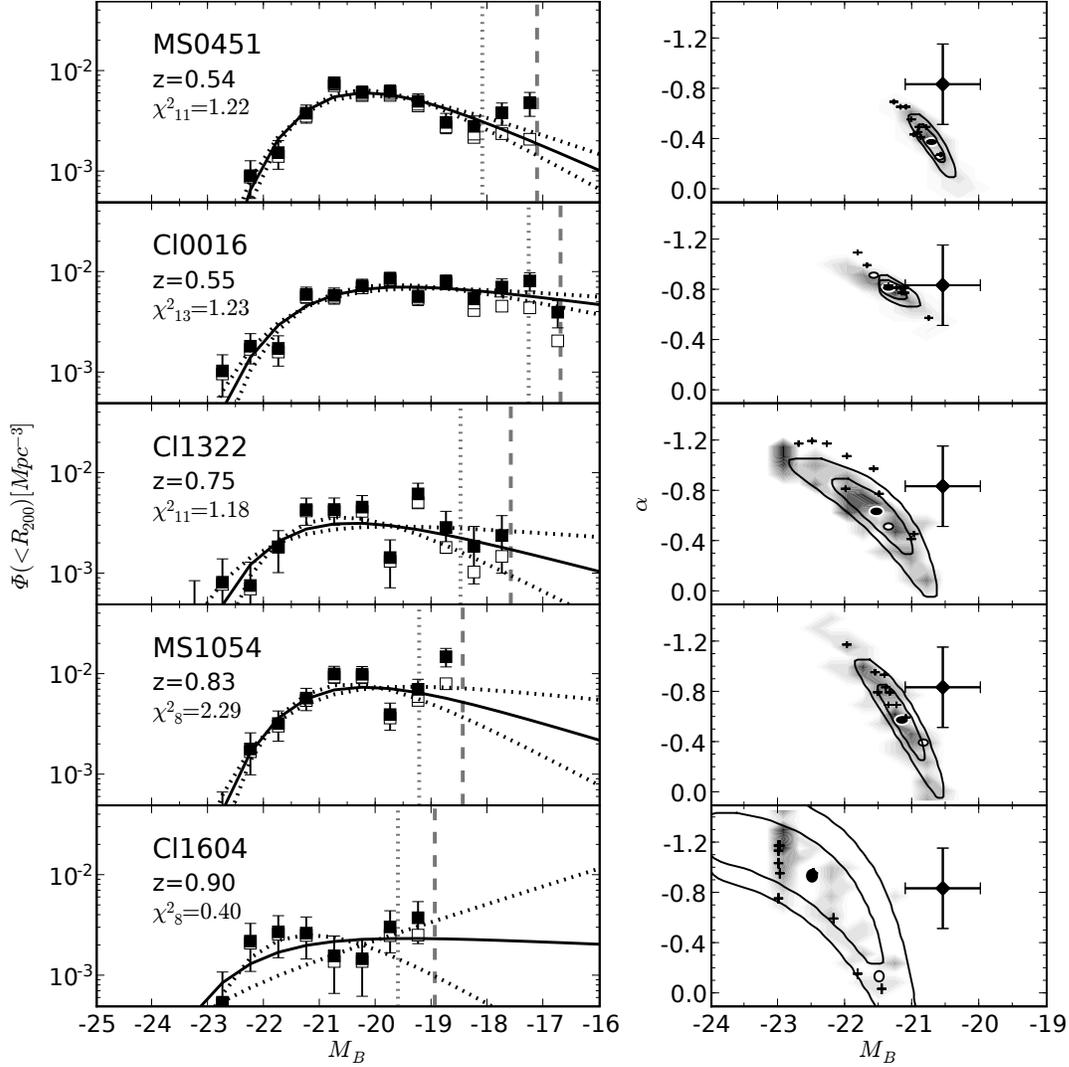}
\caption{(left) RSLF within $R_{200}$ in each of our five clusters.
  Open squares represent raw counts; solid squares are the counts
  after all corrections (see text). The best-fit luminosity function
  are solid black lines, with $68\%$ confidence-limits as gray dotted
  lines. The best-fit $\chi^2_{\nu}$ and degrees of freedom is given
  in each frame.  The 50\% and 90\% completeness limits are plotted as
  heavy dashed and dotted lines.  (right) Error ellipses ($65\%$ and
  $95\%$) are plotted for $\alpha$ and $M^*$ based on the $\chi^2$
  distribution. The best-fit value for data down to the 50\% and
  90\% detection-completeness limit are plotted as solid and open
  circles, respectively.  Best-fit values for ten realizations of the
  photometry with $\sqrt{2}$ larger errors are plotted as plus
  symbols (see text). Gray-scale represents results from jack-knife
  estimates of the errors.  We also plot the best-fit value from the
  low redshift cluster sample within $R_{200}$.  }
\label{fig:lftec}
\end{figure}

Red-sequence cluster galaxies were identified with (i) the same
color-luminosity relation (Eq. 1) as for the field, and (ii) as having
redshifts within $\pm 0.05$ of the cluster redshift. Both of these
selection criteria for our data can be seen in Figure
\ref{fig:cmzhist}. Most of the clusters do exhibit a strong red
sequence even when data are included from the entire field of view,
which is much larger than the cluster core-radius. All fields show an
over-density at the redshift of the cluster. For those two clusters
(Cl1322 at z=0.75 and Cl1604 at z=0.9) which do not show a strong red
excess in the color-redshift plot due to contrast (Figure 6, right
panel), they still show a well-populated red-sequence in the
color-magnitude diagram (Figure 6, left panel). The raw histogram in
absolute magnitude of red-sequence galaxies, normalized by the
selection volume, are displayed in Figure \ref{fig:lftec} (left panel)
for a selection radius of $R_{200}$\footnote{$R_{200}$ is calculated
based on the definition in Finn et al. (2005)}. (The volume was
estimated for each cluster from the selection in redshift and radius.)
Since the cluster is expected to cover a much narrower range in
redshift, the volume, while well-defined, is an over-estimate. Aside
from this volume normalization, this histogram represents the
uncorrected measurement of the red-sequence luminosity function in
terms of shape and luminosity normalization.

Several corrections were applied to the data to account for
observational deficiencies: 

(a) The first correction was for detection completeness, computed from
the Monte Carlo simulations (\S 2.1).  The correction is calculated
for each galaxy based on its apparent magnitude and size.

(b) Because we sample a much larger volume than the cluster, a portion
of the counts are expected to be from the field around the cluster. We
therefore subtract an appropriately volume-scaled field sample based
on the RSLF of Willmer et al. (2006), which we note is always a small
fraction of the total uncorrected cluster counts (e.g., $<10\%$ within
$1$ Mpc). 

(c) Finally, the incompleteness due to photometric redshift errors was
applied (\S 2.2), again based on simulations. Specifically, we
computed the fraction of galaxies as a function of apparent magnitude
that will be missed (scattered out of our redshift selection
window). We can ignore as negligible the fraction scattered into our
redshift selection window from the field, because to first order this
cancels with the number of field galaxies in our volume scattered out,
and further we subtract a field component from our volume, as noted
above in (b). By assuming a single spectral energy distribution for the
red galaxies, this correction can then be applied to the binned counts
as a function of absolute magnitude. 

Absolute magnitude bins containing galaxies beyond our specified
completeness limits were excluded from our calculations, and our
primary results are quoted down to 50\% detection-completeness limits.
The final, corrected data within $R_{200}$ for these limits are
presented as solid points in Figure \ref{fig:lftec} (left-hand
panel). We also use 90\% detection-completeness limits in some cases
to further demonstrate the robust nature of our statistical
results. These are indicated by the vertical dotted lines in this
Figure.

\begin{figure}[hbt]
\epsscale{0.9}
\plotone{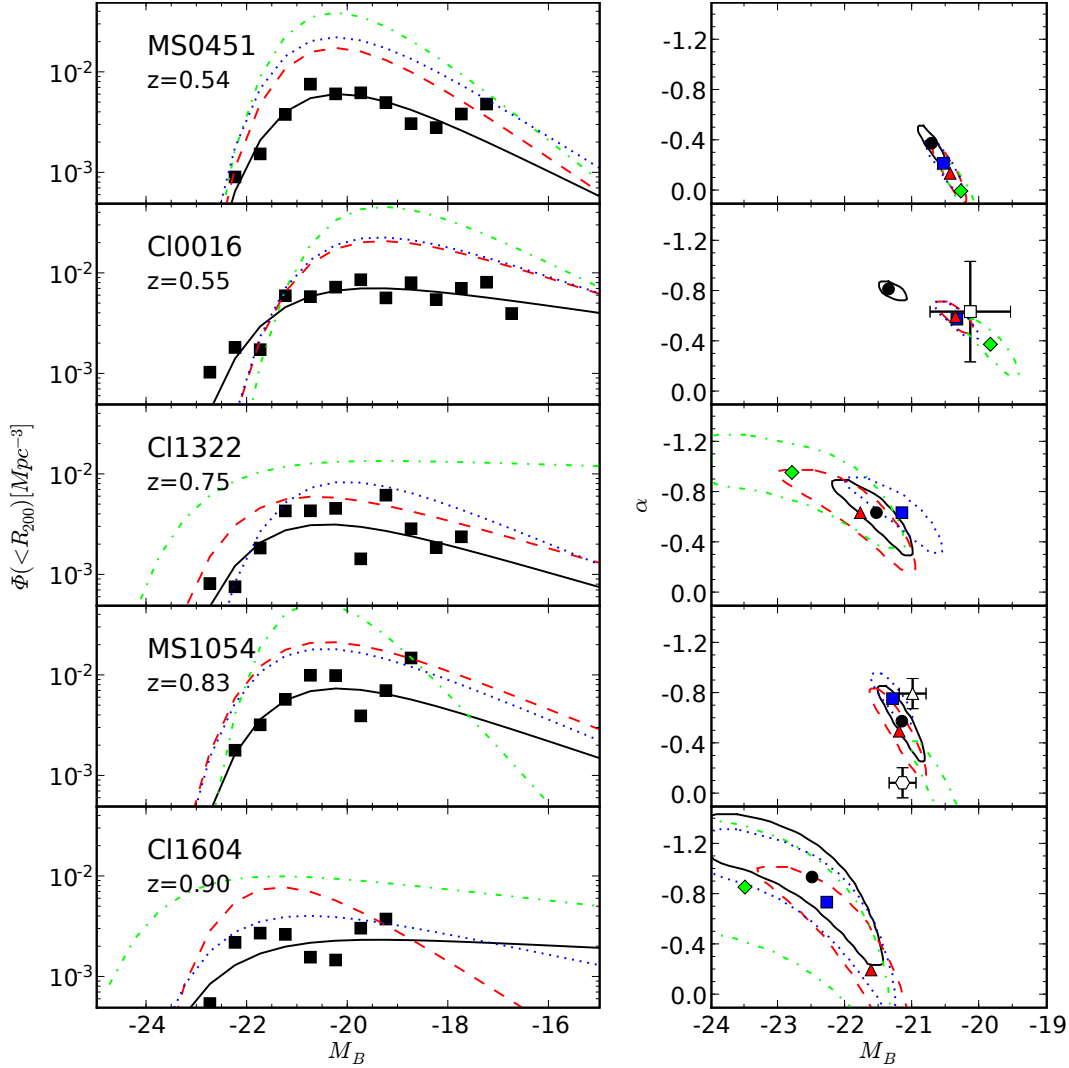}
\caption{(left) Red-Sequence cluster luminosity function for each
  intermediate-redshift cluster as measured by different selection
  radii.  Points represent corrected counts within $R_{200}$ from
  Figure 7; the solid line is the best-fit Schechter-function
  luminosity function to those points. Other lines represent best-fits
  to data extracted with different selection radii: green dot-dashed
  ($0.25 R_{200}$), red dashed ($0.5 R_{200}$), blue dotted (1
  Mpc). (right) Error ellipses ($68\%$ C.L.) for the best-fit $M^*$
  and $\alpha$ parameters of the luminosity function for all selection
  radii are shown in the left-hand panels (same line-types and
  colors).  Other measurements are shown for Cl0016 (Tanaka et
  al. 2005,square) and MS1054 (Goto et al. 2005, hexagon; Andreon
  2006, triangle). Colors are only available in the electronic
  version.}
\label{fig:lfsci}
\end{figure}

For each cluster, we calculate the best-fit Schechter-function to the
luminosity distribution by varying $\alpha$, $M^*$, and $\Phi^*$ to
find the smallest reduced $\chi^2$. Because of the nature of our
volume normalization, we focus here on $\alpha$ and $M^*$, not
$\Phi^*$, with our primary scientific emphasis being on $\alpha$. We
plot (Figure \ref{fig:lftec}, right) the error ellipses calculated
from $\chi^2$ measurements based on the best fit model. Again, these
are for data within $R_{200}$ projected about the cluster center. The
luminosity function also was calculated over several different
selection radii: $1$ Mpc, $0.5R_{200}$, and $0.25 R_{200}$. These
results and those for $R_{200}$ are shown in Figure \ref{fig:lfsci}
and listed in Table \ref{tab:lfclust}.

\subsubsection{Trends with selection radius}

Substantial differences in the shape of the luminosity function are
seen at smaller selection radii, consistent with local clusters (Lobo
et al. 1997, Popesso et al. 2006), and presumably due to a
morphology-density relation in the dwarf-giant ratio. This effect is
illustrated in our study in Figure 8.  Our clusters show a general
trend of a flatter ($\alpha \sim -1$) luminosity function with
increasing cluster radius, albeit with significant scattered
especially for Cl1322 and Cl1604, where our errors are largest.  In
the literature, Lobo et al. (1997) find a steeper faint-end slope in
the central regions of Coma as compared to groups around the
outskirts.  Popesso et al. (2006) found that the brightness of the
faint-end luminosity function increased with increasing cluster radius
in the context of double Schechter-function fits. For a single
Schechter-function fit, this would result in a flatter ($\alpha \sim
-1$) fit to the luminosity function for larger selection radius, which
is seen in our cluster sample.  RSLF shapes are far more uniform with
a $R_{200}$ selection, but field contamination becomes much greater
especially for lower-density clusters. For direct comparison with
other intermediate redshift work this result indicates it is critical
to make comparisons within the same selection radius, preferably
relative to $R_{200}$ for each cluster.  This is a point we return to
later. In general, however, we are obliged to use the $R=1$ Mpc
aperture. For our clusters, this is between 0.40-0.66 $R_{200}$.

\subsubsection{Comparison to literature}

A comparison can be made between our luminosity-function measurements
and four previous measurements for three clusters.

\noindent {\it Cl0016.} Tanaka et al. (2005) find $M^*_B=-20.14\pm0.6$
(assuming $(B-V)_0=0.9$) and $\alpha=-0.64\pm0.4$. for Cl0016 in a
selection radius that is close to $R_{200}$. Relative to our
measurement, this is $+1.2\pm0.7$ mag fainter in $M^*_B$ (a modestly
significant difference), but only $+0.2\pm0.4$ shallower in
$\alpha$. In actuality, however, Tanaka et al. use a local density
criteria for their selection. Given the well-known elongation of the
central, high-density region of Cl0016 (e.g., their Figure 6), this
results in an effective selection radius which is substantially
($\sim50$\%) smaller. Indeed, comparing to our $0.5 R_{200}$ selection
yields very similar values (see Figure 8), with their $M^*_B$ only
$+0.2\pm0.7$ mag fainter, and their $\alpha$ only $+0.042\pm0.4$
steeper. These are well within the 1$\sigma$ measurement errors, and
small in an absolute sense as well.

\noindent {\it MS1054.} Our measurements for MS1054 agree closely with
those of Andreon (2006) for his 1 Mpc selection radius:
$M^*_B=-21.0\pm0.2$ and $\alpha=-0.80\pm0.12$, which is $0.3\pm0.4$
different in $M^*_B$ and $+0.04\pm0.24$ different in $\alpha$.

However, Goto et al. (2005) find a value of $M_B^*=-21.15 \pm
0.22$\footnote{The value from Goto et al. (2005) is converted from
$i_{775}$ AB magnitudes} and $\alpha=-0.09 \pm 0.27$ for objects with
$i_{775}-z_{850} > 0.5$ within a selection radius of $R=1.35$ Mpc from
a spectroscopic sample.  Our value for $M_B^*=-21.16$ within $R_{200}$
(1.82 Mpc) is negligibly different than theirs, but we find
$\alpha=-0.58$.  As Andreon (2006) points out, the spectroscopic
completeness in Goto et al. at the faint end is only $20\%$, which may
lead to a severe turnover. Furthermore, they use a single color cut to
identify red galaxies instead of a luminosity-dependent color-cut, as
done in this work.  Consequently, bright red sequence galaxies are
likely to be included whereas faint red galaxies are likely to be
preferentially excluded due to signal-to-noise considerations and the
slope of the red sequence. If not properly accounted, this effect could
be manifest as an apparent 'deficit' of faint red galaxies compared to
bright ones.  Their color also does not span the Balmer Break at
$z=0.83$.

As a further check of the Goto et al. measurements, we note the
luminosity function they derive for {\it early-type} galaxies is
inconsistent with their measurement for {\it red} galaxies. Early-type
galaxies are selected in their study via a visual (qualitative)
morphological classification based on HST images. On the other hand,
their value of $\alpha = -0.54 \pm 0.13$ for early-types galaxies is
very comparable to our measurement.  The early-type galaxies at this
redshift are still overwhelmingly red (and dominate the counts of red
galaxies in MS1054; van Dokkum et al. 1999). No population of bright,
red, late-type galaxies exist in MS1054 that would bias the luminosity
function in such a manner as might be inferred by the Goto et
al. measurements.  Based on the luminosity-function agreement between
studies, we conclude their morphological selection is more reliable,
in this case, than their color selection.

\noindent {\it Cl1604.} Andreon (2008) measured the luminosity
function within the central 0.45 Mpc of the cluster from two-band HST
imaging.  He derived a value of $\alpha=-0.67\pm0.33$ for the slope of
the faint end.  Our value of $\alpha=-0.86\pm0.46$ within an aperture
of R=0.37 Mpc agrees within the $1\sigma$ limits.

\subsection{Reliability of the Intermediate-Redshift RSLF Determination \label{sec:lftest}}

To further verify the accuracy of our results, we performed a number
of tests of our data, reported here, including four tests of the
measurement reliability given the known amplitude of random errors,
two tests investigating the systematic impact of the selection
function for red galaxies, and a final test of the surface brightness
bias present in our sample. Each of these tests validate a different
aspect of our cluster measurement and reveal the quality of our data
and analysis, and the robustness of our results on the RSLF
measurement.

\subsubsection{Random-Error Effects}

The first test we conducted was to measure the completeness and
reliability of detecting red galaxies within the cluster volumes
selected.  We calculated the colors of a red galaxy (from synthetic
spectra) at the redshift of each cluster, and at a redshift $\pm$ 0.1
and 0.15 about the cluster redshift, i.e., at 2 and 3 times the
distance in redshift as our nominal selection cut ($\pm$ 0.05 about
the cluster redshift). We then simulated the appropriate multi-band
magnitudes and errors for the object for $R=19$ to 26, in steps of 0.5
mags. For every magnitude interval and redshift bin, the simulated
galaxy's photometry was realized 100-500 times as a perturbation about
the nominal colors, drawing statistically on the error distribution
for each band to determine the perturbation. For each realization the
photometric redshift was calculated. To determine completeness we
simply counted the fraction of the galaxies simulated at the cluster
redshift that retained photometric redshifts within $\pm$ 0.05 of the
cluster redshift. To determine the reliability, we calculated the
percentage of galaxies simulated at $\pm$ 0.1 and 0.15 away from the
cluster redshift had photometric redshifts estimated to lie within
$\pm$ 0.05 of the cluster redshift. Such galaxies would be selected in
our scheme as red-sequence cluster galaxies. This percentage was then
normalized by the ratio of the number of galaxies in the cluster
vs. the number of galaxies in the volume from $z_{cluster}\pm 0.15$,
as estimated in the real data. The contribution from the more distant
shell was vanishingly small compared to the nearer shell, indicating
our simulation should be quite accurate. 

The results of this simulation are shown in Figure 9. From this it is
clear that contamination is always $<$10\%, and typically only a few
percent even at the 50\% source-detection limit. The selection
completeness is $0.78\pm0.15$\% at the 50\% detection limit, and
$0.86\pm0.13$\% at the 90\% detection limit. This is roughly what one
would expect given the S/N and associated photometric-redshift errors
at these limits. If anything, the results are somewhat optimistic,
which can be understood in terms of the idealized nature of the
simulation, e.g., the simulated galaxies are drawn from the same set
that are used to detect and derive photometric redshifts.  Overall,
however, the simulation shows we are able to recover close to the
expected number of cluster sources in a controlled situation
resembling the actual data under analysis.

The remaining three tests directly probe the derived luminosity
function itself.

In the second test, we exclude galaxies from our RSLF calculation
where the detection-completeness was less than $90\%$ (instead of the
$50\%$). While we expect using corrected data down to the $50\%$
completeness-limit is reliable (due to the extensive completeness
simulations performed and the corrections derived therefrom), the
robustness of our results is most directly shown by examining the
truncated data set. We follow the same steps to calculate the
luminosity function parameters as described in \S 3 but with fewer
points. As can be seen by comparing the open to filled diamonds in
Figure \ref{fig:lftec} (right), none of the clusters show a
significant change in the parameterization of the luminosity function.
At $90\%$ completeness, all of our data extends beyond $M^*+2$.

The third test was to create 100
realizations of each cluster RSLF from the measured errors in the
corrected counts of galaxies identified as red-sequence cluster
members.  For each cluster, the observed luminosity function was
convolved with the errors at each magnitude.  Then, the
parameterization of the luminosity function was measured for each
realization, and the averages for the Monte-Carlo simulation were
computed.  The values found for $M^*$ and $\alpha$ are plotted as
gray-scale in the left-hand plots in Figure \ref{fig:lftec}.  For all
the clusters, the averages are well within the $1\sigma$ error
measurements for the parameterization of the cluster luminosity
function and the variance is of the same order as well.

The fourth and final test was a more extensive test of the entire
process of measuring the cluster luminosity function. Following the
procedure of Toft et al. (2004), we produced ten realizations of our
photometric catalogs for each cluster. In these realizations, the
measured aperture photometry was smeared, in a statistical fashion,
drawing from the photometric error distribution for each flux
measurement. This means that the error distribution for each
realization is roughly $\sqrt{2}$ larger than the initial (measured)
catalog, although we did not update the effective error distribution
for these realizations.  With the smeared photometry, we recalculated
the photometric redshifts. These new redshifts were then applied to
the calculation of rest-frame properties and the selection of cluster
galaxies. The luminosity functions were built using the same
corrections and procedures as previously described. Finally the
parameterization of the luminosity function was measured for all of
the clusters. The results of each of the realizations are presented in
Figure \ref{fig:lftec}. For all but one of the clusters, the
realizations produce results which show no significant difference in
the mean from the measurement of the luminosity function. The
dispersion in the luminosity function parameters, however, is larger,
as expected from the additional noise introduced in the simulation
process. For one of the smallest cluster in our sample (Cl1322), the
results indicate a {\it flatter} ($\alpha \sim -1$) luminosity
function then measured in the single data alone. However, the original
measurement is contained within the spread of slopes that are found,
and is not statistically significant.

\subsubsection{Systematics with Color-Selection}

To test what impact the specific selection of red cluster galaxies has
on the derived RSLF, we selected galaxies from the ten
realizations of each cluster catalog by applying perturbations to the
color, magnitude, redshift, and radial selection functions.  

Changes to the magnitude zero-point of the color-magnitude relation
made no changes to the selection of red galaxies due to the steep
nature of the relationship.  Large changes ($>0.1$ mag) to the color
zero-point did result in a small shift in the faint end slope of the
luminosity function, with a tendency to find steeper downturns (more
positive $\alpha$ by $+0.15$) with a redward shift in the cutoff, and
flatter slopes (more negative $\alpha$ by $-0.14$) with a blueward
shift. This is qualitatively consistent with the canonical picture
that the red-galaxy population and blue-galaxy population are
characterized, respectively, by shallow and steep faint-end slopes to
their luminosity functions. A similar trend was observed with changing
the slope of the color-magnitude relationship.  However, neither
change to the color selection-function would result in a measurement
outside of the $95\%$ confidence limits for the value found for the
original color selection. This result is in qualitative agreement with
previous analysis (Andreon et al. 2006, de Lucia et al. 2007,
Barkhouse et al. 2007).

\subsubsection{Systematics with Redshift Window and Radius}

Variation in the redshift window do result in a small systematic
change in the luminosity function. The magnitude of the change,
however, is much smaller than the errors on the measurement of an
individual cluster: $\Delta \alpha < \pm 0.1$ for a factor of two
change in the redshift window.  When the redshift window is increased,
the luminosity function becomes steeper (closer to a flat faint-end
with $\alpha \sim -1 $), and becomes shallower (closer to $\alpha \sim
0$) when the window is decreased.

The systematic change in the luminosity function with the change in
selection window may indicate an underestimation of the
photometric-redshift error-correction or contamination from field
sources from the larger volume being investigated.  Without extensive
spectroscopic redshifts at faint magnitudes, it is difficult to
conclude the source of this bias.  Alternatively, this may be similar
to the general trend seen in the change of slope with selection
radius, albeit washed out by errors in photometric redshift, i.e., a
steepening of the slope as the core is more preferentially sampled.
Regardless, the small magnitude of the change provides confidence in
our measurement and the robustness of our corrections.

Small changes to the radial selection function do not result in
significant changes to the luminosity function. Trends with larger
variations in the radial selection, already noted, will be discussed
again later.

\subsubsection{Surface-Brigtness Selection Effects}

Finally, we check that the depth of our observations -- in terms of
surface-brightness sensitivity -- are sufficient to detect cluster
members at these redshifts. As can be seen in Figure \ref{fig:sbias},
the size-magnitude locus of the red sequence sources for most of the
clusters is well within our surface-brightness detection limits. Only
for the highest redshift cluster, Cl1604, are we truly in danger of
missing some of the objects. 

However, if we are missing a significant amount of objects at the
faint end, this will have the effect of causing the slope to fall more
steeply than it should ($\alpha$ too large), which would mimic the
astrophysical effect claimed by others to be an evolutionary
phenomenon. We do not see this effect in Cl1604 or MS1054, which both
have slopes on par with our other clusters, and in fact Cl1604 has
nominally the most negative $\alpha$ in our sample.

\subsection{The RSLF at $z \sim 0.1$ \label{sec:lowz}}

A number of studies have measured the $z\sim0$ luminosity function
from a variety of different sources using a wide range of techniques.
To provide a single, simple comparison to our body of work, we have
measured the RSLF for a large sample of Abell clusters (Abell, Corwin,
\& Olowin 1989) based on SDSS imaging data and spectroscopy
(Adelman-McCarthy et al. 2008) using a similar procedure as for our
clusters.  Clusters are selected from the Abell catalog that also
appear in the SDSS DR6 imaging and spectroscopic catalogs.  For all
the clusters that do appear in the SDSS data, we measure the velocity
dispersions of the clusters based on the spectroscopic data.  From
these measurements, we only include clusters with a minimum of 30
spectroscopically confirmed members.  Clusters with extremely high
velocity dispersions ($\sigma_v > 1500$ km s$^{-1}$) or with significant
difference in richness and $\sigma_v$ are examined individually to
confirm that they are not close superpositions of two smaller
cluster. Superpositions are eliminated from the sample. Finally,
cluster velocity dispersions are compared to values found in the
literature from Struble \& Rood (1991), Wu, Xue, \& Fang (1999),
Miller et al. (2005), or Popesso et al. (2005).  Clusters with large
disparities between their literature value and that measured by the
SDSS data were removed as well.  Our final sample of 59 clusters have
redshifts between $0.035 < z < 0.144$ and $\sigma_v > 500$ km s$^{-1}$
with average values of $z=0.078$ and $\sigma_v=886$ km
s$^{-1}$. Twenty-six of the clusters have $\sigma_v>900$ km
s$^{-1}$. The selected clusters are listed in Table \ref{tab:lowz}.

Using the SDSS photometric data, we have identified and analyzed the
data following the same procedure as our cluster sample.  The primary
difference is that we only use the five SDSS bands that are available
and no narrow band data.  For magnitudes, we used corrected Petrosian
magnitudes following the recipe in Graham et al. (2005), and for
colors, we use the SDSS fiber magnitudes, which are analogous to our
aperture magnitudes.  We have employed the same photometric redshift
technique with a training sample created from the SDSS spectroscopic
sample, calculated absolute magnitudes and rest-frame colors, and
selected red sequence galaxies using the same selection function
except shifted to z=0.08 to account for luminosity evolution.  Cluster
galaxies were selected by having photometric redshifts within $0.05$
of the cluster redshift.  The luminosity function was calculated in
the same manner as for our clusters with corrections applied for
photometric redshift errors and field subtraction. However, since the
magnitude limit of the SDSS at low redshift probes much deeper down
the cluster luminosity function, no correction for incompleteness need
be applied for measuring the luminosity function to $M_B^*+3$.  The
luminosity function for each cluster is reported in Table
\ref{tab:lowz} within a selection radius of $R_{200}$.

A large dispersion in the value of $M_B^*$ and $\alpha$ is present for
the clusters.  Averaging all of the clusters together, we find
$M_B^*=-20.55\pm 0.56$ and $\alpha=-0.84 \pm 0.32$ within $R_{200}$,
which is a significantly larger dispersion than if we had first summed
the clusters together and then measured the luminosity function (see
Appendix \ref{app:cum} for further issues with ensemble-averaged
luminosity functions) or from the measurement error associated with an
individual cluster (typically around $\sigma_{\alpha}\sim0.15$).  For
$R=1$ Mpc, we find values of $M_B^*=-20.39\pm 0.48$ and $\alpha=-0.71
\pm 0.32$. These values are very comparable to similar studies once
converted to our magnitude system.  Figure \ref{fig:zev} and Table
\ref{tab:survey} contain values and references to other measurements
of $M_B^*$ and $\alpha$ for low redshift clusters.

\section{Evolution of the RSLF \label{sec:evo}}

The central question of this is work is determining changes in the
cluster RSLF shape with redshift. To this end, the luminosity function
parameters, $M_B^*$ and $\alpha$, are plotted in Figure \ref{fig:zev}
for individual clusters in our low- and intermediate-redshift samples,
their mean values binned in redshift, and other clusters' values
published in the literature. The latter have been transformed into
our magnitude system.  Measurements of the luminosity function based
on isolating the red-sequence via morphology, single-function fits to
deep luminosity functions ($M^*+5$; see \S 3.3 and 5.2), and in galaxy
groups have been excluded. To provide the closest comparison to other
studies, we plot the data for a cluster radius of 1 Mpc.  Values for
the parameterizations within $R_{200}$ instead of 1 Mpc generally are
closer to $\alpha=-1$ with a brighter $M^*$ for both the low and
intermediate redshift clusters in our sample, but give qualitatively
the same trends with redshift.

Our clusters exhibit an increase in $M_B^*$ with redshift, as also
seen for the field RSLF (Willmer et al. 2006). However, the cluster
RSLF $M^*$ is 0.5-1 mag brighter in the field value at $0.7<z<0.9$,
but 0.5 mag fainter at $z\sim0.5$--i.e., brightening with redshift
appears steeper in clusters for $z>0.5.$ In contrast, the value of
cluster RSLF $M^*$ is relatively flat between $z=0-0.5$, considering
in concert local values from our low redshift sample or the
literature. Overall, the brightening of rest-frame $B$-band $M^*$ in
clusters and the field agree well with the findings by de Propris et
al. (1999) and Lin et al. (2006) in the near-infrared for the fading
of simple stellar populations formed between $1.5<z<3$.

We find no trend in the faint end slope, $\alpha$, with redshift
within our sample.  For the low redshift clusters, $\alpha$ shows a
large range of values.  However, the average value for the low
redshift clusters within $R=1$ Mpc is $\alpha = -0.68\pm0.42$, which
increases to $-0.71\pm0.29$ if we only include clusters with $\sigma_v
> 1000$ km s$^{-1}$. Using this number for the high mass clusters
(which is probably most comparable to our cluster sample), only one
cluster, MS0451, is significantly different at the $1.7\sigma$ level.
Including this cluster, the intermediate redshift sample has an
average of $\alpha = -0.59 \pm 0.26$, which is not significantly
different than the average value for the low-redshift cluster sample.

To further investigate the question of evolution, we plot $\alpha$ as
a function of cluster velocity dispersion for our full sample of low
and intermediate redshift clusters in Figure 13. The data are
calculated within a cluster radius of $1 Mpc$ and $R_{200}$.  The mean
of the two distributions is different but not at a significant
level. From our sample of intermediate redshift clusters, we find no
significant evidence for a deficit of galaxies occurring to $M^*+3$ as
compared to low redshift clusters.

\section{Discussion}

\subsection{Assessment of Results in the Literature}

Recent literature contains claims and refutations of a deficit of
faint, red galaxies in intermediate redshift clusters. A deficit would
be significant because a change in the RSLF {\it shape} implies an
ongoing or multi-epoch formation scenario, beyond passive evolution of
a coeval population forming at high redshift. We suspect different
conclusions regarding the deficit (or lack thereof) arise in part from
differences in samples, analysis methods, or comparisons to local
samples. We enumerate these points below. For comparison and reference
purposes during this discussion, we list all of the relevant studies
of the red sequence cluster luminosity function and their principal
attributes in Table \ref{tab:survey}.

\subsubsection{Analysis methods \label{sec:dgr}}

A number of studies measure the luminosity function shape in terms of
a dwarf-to-giant ratio (DGR). Typically, the DGR is defined as a ratio
between the number of galaxies in two magnitude bins, and as such, it
is a much simpler calculation than the luminosity function.  However
the DGR suffers from a number of problems rarely addressed in the
literature. First and foremost, the DGR has had a range of definitions
({\it cf.} Ferguson \& Sandage 1991, Secker \& Harris 1996, Driver,
Couch, \& Phillipps 1998, Tanaka et al. 2005, de Lucia et al. 2006,
Gilbank et al. 2007, Koyama et al. 2007, Stott et al. 2007 ), and
consequently is not directly comparable between all studies. Often the
DGR is based around observational (i.e., detection) limits such that
the definitions of ``dwarf'' and ``giants'' do not necessarily reflect
any natural division between galaxies and ignores traditional splits
between such systems. A further weakness of the DGR is that its value
is dependent on accurate measurements from each of two magnitude bins,
whereas the luminosity function (as shown in Section \ref{sec:lftest})
is relatively insensitive to variation in a single bin. As the
faintest magnitude bin is likely to have the largest uncertainties due
to incompleteness, photometric errors, redshift errors, or other
sources, an accurate statistical calculation of the DGR has far
greater errors than a similar value derived from the fit of the
luminosity function.

One of the purported strengths of the DGR is that it does not suffer
from the covariance between $M^*$ and $\alpha$ that is a common
complaint of Schechter-function fitting.  However, without measuring
the shape of the luminosity function, it is difficult to ascertain
whether changes in the DGR are due to changes in the bright end
(dominated by $M^*$) or changes in the faint end (dominated by
$\alpha$). While the DGR is strongly correlated to changes in
$\alpha$, it is also weakly correlated with changes in $M^*$. We
illustrate this in Figure \ref{fig:dgralpha}.  Following the
definition of de Lucia et al. (2006) for the DGR: \begin{equation} DGR
= N(M_V < -20.0) / N(-20.0 < M_V < -18.2), \end{equation} it can be
shown the DGR will vary approximately as $log_{10} (DGR) \sim
-0.67\alpha$ for constant $M^*$ and $log_{10}(DGR) \sim 0.62 M^*$ for
constant $\alpha$.  Variation in the DGR therefore may not only be due
to evolution in $\alpha$.  As will be discussed later (also see
Appendix A for issues with ensemble-averaged luminosity functions)
cluster-to-cluster variation and sample selection may affect the
measurement of the DGR to bias the results and confuse variation in
the bright-end normalization for evolution in the faint end slope. For
these reasons we prefer a parametric estimate of the luminosity
function via the Schechter fitting-function.

\subsubsection{Survey quality}

Significant variations exist between surveys in terms of the level of
analysis and the quality of the data. For example, a significant
difference between our study and most RSLF studies in the literature
is the amount of simulations undertaken to understand the completeness
and selection functions of the data. Few other studies have undertaken
extensive analysis of their incompleteness, with several notable
exceptions. For example, Barkhouse et al. (2007) undertook a number of
simulations outlined in their appendix to understand the effects of
projection, Eddington Bias, and color selection.  Mercurio et
al. (2006) performed extensive completeness simulations for analysis
of a low redshift cluster, but then adopted conservative luminosity
limits well above their detection- and selection-completeness
limits. Andreon (2006) presented an in-depth description of the
statistical method for calculating the luminosity function in the
limits of a background population. However, of these examples, the
first two focus on low-redshift clusters. Only Andreon's (2006) study
is at comparable intermediate-redshifts as our own, and, notably,
finds similar results as our study.

Imaging depth also varies substantial from survey to survey. None of
the intermediate redshift studies explore beyond $M^*+3.5$, and many
are substantially (1-2 mag) shallower, in contrast to many of the
low-redshift studies, Shallow surveys are not able to probe the faint
end of the luminosity function, and are really just probing the giant
population.

Another salient difference between surveys is wavelength coverage,
and the impact this coverage has on reliable selection of red-sequence
galaxies.  Our study, although smaller in the number of clusters than
many of the other studies, has far greater wavelength coverage and
information to constrain and identify the cluster populations.  Many
surveys use only two bands and background-subtraction to identify
cluster sources. In contrast, it has been shown that multi-band data
sufficient to construct precision photometric redshifts can be far
more reliable in selecting cluster galaxies (Brunner \& Lubin 2000,
Rines \& Geller 2008).

\subsubsection{Band-pass effects}

As has been shown by Goto et al. (2002), the slope of the RSLF is
passband-dependent.  We have avoided comparing our results to any of
the papers that measure the K-band luminosity function, but the
conversions between other optical bands may still suffer from issues
beyond simple color transformations.  Indeed, further complications
between measuring the RSLF in different passbands was initially shown
by Smail et al. (1998).  When measuring the apparent I-band luminosity
function for galaxies in $z\sim0.25$ clusters, a selection in apparent
(U-B) resulted in a deficit of galaxies as compared to the same sample
but selected in (B-I), which covered the 4000 $\AA$ break at that
redshift.  To maintain consistency, Andreon (2008) always used filters
saddling the 4000 $\AA$ break regardless of cluster redshift.
However, a number of studies (e.g., de Lucia et al. 2006, Gilbank et
al. 2008) have not done this.

\subsubsection{Radius effects mixed with sample selection}

A further difficulty in comparing the luminosity functions from
different studies is the selection radius adopted in analysis. The
luminosity function does show variations with cluster radius (Popesso
et al. 2006, Barkhouse et al. 2007, and \S 3.1.1 here); a selection of
a small, fixed radius will bias results for clusters over a range of
masses.  This affect can be especially pronounced when comparing
sources at different redshifts.  The Stott et al. (2007) sample is an
example of the possible problems that may arise when using a small
selection radius of fixed metric size. The low redshift clusters in
their sample have smaller mass (in their case, they use x-ray
luminosity as a mass proxy) than the higher-redshift sample.  However,
they use a constant measurement radius of 0.6 Mpc for all of their
clusters.  For the intermediate redshift clusters, they are only
measuring the core of the cluster where the decline in the faint-end
of the luminosity function tends to be more pronounced, but for the
low redshift clusters, they measure out to $R_{200}$. Unsurprisingly,
they measure a deficit for the higher-redshift clusters. Indeed, for
the one low redshift cluster of similar x-ray luminosity as the
intermediate redshift clusters (Abell 3555 in their sample with a
$DGR=1.89$), they find a DGR similar to the intermediate redshift
clusters. In contrast, Andreon (2008) also uses a nearly constant
radius of 0.5 Mpc for the high redshift clusters in his sample, and he
finds no trends with redshift, but fairly significant scatter between
clusters.

\subsubsection{Sample differences}

Our selection of clusters samples the most massive structures in the
Universe -- clusters that are almost of factor of two more massive
than the EDisCS sample (de Lucia et al. 2006), as shown in Figure
\ref{fig:zmass}. According to the simulations of Wechsler et
al. (2002), the typical line of sight velocity dispersion of our
clusters, today, should be around $\sigma = 1500$ km s$^{-1}$, placing
their masses well above the Coma cluster. None of the local comparison
surveys, or even the CNOC cluster sample (Yee et al. 1996) at lower
redshift are a fair match in mass to our sample; the large sample of
Popesso et al. (2004, 2005, 2006) with velocity dispersions estimated
from their x-ray luminosity are also far less massive for the most
part.  The closest comparison is the Smail et al. (1998) sample with a
few clusters overlapping our sample, but with a much lower average
mass.  The average mass of the Stott et al. (2007) intermediate mass
sample is only slightly less than our sample, but their low redshift
sample is far less massive.  (We estimate the velocity dispersion for
the Stott et al. sample either from their x-ray luminosities, or from
the velocity-dispersion listed in Andreon (2008) for the overlapping
subset.)  In Andreon (2008), the very high redshift clusters are, on
average, a factor of 30\% less massive than the $z\sim0.55$ sample.
The masses of clusters in the Barkhouse et al. (2007) and Gilbank et
al. (2007) studies are estimated using the $B_{gc}$ parameter, which
is a measure of the cluster richness. As they mention, this estimator
may have dependencies related to redshift. According to their
conversions, most of their sample is far less massive than those
studied here.  Our low-redshift clusters do span a large mass range,
but a third of the clusters do have $\sigma_v > 1000$ km s$^{-1}$,
which is comparable to the predicted masses of our cluster sample. If
luminosity function shape, or its evolution, depends on cluster mass,
than cluster sample selection could be responsible for the different
results found in the literature.

For illustration of possible sample effects, we plot in the bottom
panel of Figure \ref{fig:zev} the amount of evolution in $\alpha$ (the
faint-end slope) seen by Stott et al. (2007), assuming the value for
$\alpha$ of our low redshift clusters is correct.  To make this
comparison, we have converted their measure of the DGR evolution (a
power-law dependence in $1+z$) to a function of $\alpha$ (see Figure
11) using the relationship found between DGR and $\alpha$ in \S
\ref{sec:dgr} and assuming constant $M^*$. As can be seen, this trend
is inconsistent with our data, even when normalizing to our
low-redshift data. Fitting to our own data, we find a much shallower
trend which is consistent with no change.  It is conceivable, we
suggest, the difference in redshift-trends are due to differences in
the two samples' cluster masses, and systematic trends in cluster-mass
with redshift in the Stott et al. sample.

In support of this argument, Koyama et al. (2007) find a trend between
cluster mass and $\alpha$ in the sense that more massive clusters have
steeper turnovers.  We see a similar trend in our intermediate
redshift clusters (Figure 13). However, the strength of the trend
appears sensitive to the radial selection, and our statistics are
poor. In contrast, there is no apparent trend in our low redshift
data. Andreon (2008) also sees no trend using a larger sample of
clusters. On balance, while cluster-mass may play a role in explaining
differences between survey results on the RSLF, it is more likely that
cluster-mass differences {\it within} samples, particularly when
correlated with redshift, plays a more substantial role in driving
apparent evolutionary trends seen in some studies.

\subsection{Challenges to Measuring Evolution}

In the previous section our focus was on identifying survey
differences that could explain, in part, the discrepant results found
in the literature on RSLF evolution. Here we turn instead to the
astrophysical nature of the RSLF, and how its complexity and variation
presents fundamental challenges to its measurement.


In the low redshift Universe, an over-abundance of ``true'' dwarf,
early-type galaxies (those with luminosities below $M_B^* < -17$) are
found in clusters compared to the field (Driver et al. 1994, de
Propris et al. 2003) and lower density environments (Trentham et
al. 2005), indicating some relationship between the environment and
shape of the faint RSLF must exist. In fact, since the RSLF is a
composite population (e.g., massive ellipticals, intermediate mass
lenticulars, and low-mass spheroidals) a single Schechter function is
likely an inadequate description of the RSLF; bumps and dips have long
been noticed in cluster luminosity functions (Kashikawa et al. 1995,
Jerjen \& Tammann 1996). Since these different sub-populations of the
red-sequence have different relative densities with environment,
changes in the shape of the RSLF with environment is to be expected at
all luminosities. This is likely what drives the observation that
cluster RSLFs changes shape with selection radius, and perhaps also
differences between clusters.


Due to the composite nature of the RSLF, a two-function fit (either
two Schechter functions or a Gaussian for the luminous component plus
Schechter) is commonly used for parameterizing low-redshift cluster
luminosity-distributions (Biviano et al. 1995, Popesso et al. 2006).
When only a single function is fit to the luminosity function, the
faint end slope is strongly affected by the limiting magnitude. In
intermediate-redshift cluster studies, single-function fits are made
exclusively, yet as Table 4 shows, there is substantial variation in
sampled depth.

Since the SDSS data has much greater depth then our
intermediate-redshift sample, we can explore the amplitude of the
effect of measuring the RSLF to different depths for a given selection
radius. At $M_B^*+5$ the faint-end slope should be dominated by the
dwarf population (Trentham \& Tully 2002, Popesso et al. 2006). If we
measure $\alpha$ to $M_B^*+5$ instead of $M_B^*+3$, we find a faint
end slope of $\alpha = -1.15$ instead of -0.84, i.e., the luminosity
function appears to be slightly rising instead of falling at these
greater depths. (This value is possibly an underestimate as
incompleteness will start to effect the number counts at $M_B^*+5$ in
the SDSS data; no correction for incompleteness has been applied to
these data.) In other words, as the sampled depth decreases relative
to M$^*$ (as we would expect and indeed has happened in higher
redshift surveys), the apparent slope of the RSLF increases, i.e.,
an apparent deficit appears.

The reason for this effect is simply that the bright-end of the RSLF
is typically characterized in shape as a `bump,' i.e., having a
maximum near M$^*$. This bright bump is dominated by giant galaxies.
The apparent decline at lower luminosities is substantially modulated
by the relative amplitude of the dwarf-galaxy luminosity function, but
intermediate redshift surveys do not reach down to faint enough
magnitudes to well characterize the luminosity function of this
population itself. Our data for MS0451 is a good example. In this case
there is some evidence for an up-turn in the RSLF at the faintest
magnitudes and a dip occurring prior to this upturn about 2-2.5 mag
fainter than M$^*$, indicative of a composite population. However, the
faint-end slope of the luminosity function we measure down to M$^* +
3.5$ is dominated by the excess of bright galaxies.


Survey-depth alone is not a panacea.  The above conclusions are
modulated by variations seen between clusters RSLFs and possible
correlations with mass.  There are indications that the RSLF M$^*$ is
brighter in clusters than in the field (de Propris et al. 2003, but
Hilton et al. 2005 notes a few exceptions), and brighter still in more
massive clusters (Croton et al. 2005, Hansen et al. 2005).  Similarly,
for a sample of 10 clusters, Yagi et al. (2002) find the dip in the
luminosity function around $M_B\sim-18$ becomes much stronger in
higher-mass systems.

Unfortunately, there is considerable scatter that washes out any clear
correlation between the shape and normalization of the luminosity
function and cluster mass for large samples (Biviano et al. 1995,
Popesso et al. 2003, de Propris et al. 2003 and 2005, Barkhouse et
al. 2007). Within our low redshift sample, $M_B^*$ does brighten with
cluster mass within $R_{200}$, but the correlation is not
statistically significant.  Variations in the complex
luminosity-function shape are seen in the nearby clusters of A168
(Yang et al. 2004) and Shapely super cluster (Mercurio et
al. 2005). In our survey, variations are most obvious in comparing two
of our $z\sim0.55$ clusters.  Even though both are at the same
redshift and are of similar mass, MS0451 has a very pronounced bright
end as compared to the relatively flat bright end of Cl0016.

Cluster to cluster variation is a significant effect and a detriment
to measuring evolution from a small, diverse sample of clusters. The
alternative, namely, to studying large samples of clusters to average
over variations also presents a challenge, since each cluster must be
observed to adequate depth to well characterize the
luminosity-function, as discussed above.

Along these lines we close with one further caution pertaining to
large samples. With significant variation in the luminosity function,
the ensemble average luminosity distribution may not be the same as
the average of the individual luminosity distributions.  Simulations
we have performed (see Appendix \ref{app:cum}) indicate systematic
errors do result from ensemble averaging, and are particularly large
for the DGR index. We conclude that in order to confirm the hypothesis
of evolution of the red sequence, a far more significant sample of
intermediate redshift clusters has to be observed to greater depth to
constrain the true dwarf population.  These observations must be
analyzed in a manner consistent with analysis of low-redshift cluster
samples, and should not be ensemble-averaged.

\subsection{Evolution Scenarios}

If we take the results of de Lucia et al. (2006) and Stott et
al. (2007) at face value, in concert with ours, we conclude the red
sequence is already in place at intermediate redshifts down to faint
magnitudes ($M_B \sim -18$) in the most massive clusters, whereas the
build-up is still occurring in lower-mass clusters.  Mass-dependent
evolution would be in concordance with a general trend of down-sizing
in hierarchical structure-formation scenarios, where the most massive
structures form earliest. Here, this would have to occur on two
scales: (i) The most massive clusters would be the ones to have their
red sequences in place first. (ii) For any given cluster, the most
massive galaxies would evolve to the red sequence first.

Another more subtle possibility may be occurring. While the
hierarchical structure-formation scenario addresses directly the
assembly and build-up of mass, there is the associated notion that the
pace of star-formation is also set by this build up.  Specifically we
expect to see, and perhaps do see, a down-sizing in the co-moving
star-formation rate such that it is dominated by lower-mass systems at
later times.  Star-formation, however, is stochastic on a galaxy
scale. Further, small-mass bursts in large-mass systems can
dramatically change the colors, providing a blue, if ephemeral,
photometric icing. Consequently, while the mass build-up may indeed be
hierarchical, the growth of the RSLF as a function of luminosity may
be substantially modulated by on-going star-formation. If this is the
case, the evolution of the {\it shape} of the RSLF (as distinguished
from the {\it normalization}) may be a rather subtle phenomenon, as we
observe within our own study.

By looking at the star-formation histories of galaxies currently on
the red sequence, we can put some additional constraints on the RSLF
evolution. Studies place the formation epoch of red-sequence galaxies,
(particularly the massive ellipticals) at $z>2$, with only passive
evolution since that time (e.g., Holden et al. 2004). However,
elliptical galaxy stellar populations are typically found to be far
more uniform than S0 populations, yet S0's are counted in the RSLF,
and are a significant -- if not dominant -- component at bright
absolute magnitudes in local samples. If a deficit of low-luminosity,
passively evolving galaxies existed at intermediate redshifts, then we
would expect to see some correlation between stellar ages and
magnitude in local cluster red-galaxy populations. Perhaps S0's are
part of this picture.

Indeed, some S0 galaxies are relatively recent additions to the
cluster environment.  They are missing in intermediate clusters
relative to local clusters (Dressler et al. 1997), and almost $40\%$
of those that are found at intermediate redshifts have spectra
indicative of recent star formation (Tran et al. 2007).  In some local
clusters, S0's show far greater spread in their stellar histories than
elliptical populations (Kuntschner \& Davies 1998), and the
color-magnitude relationship is far tighter for ellipticals than S0
galaxies (Bower et al. 1992).

If an evolving S0 population is to be responsible for a deficit in the
intermediate redshift RSLF at faint magnitudes, critical then is
determining the luminosity range over which the S0's dominate the
RSLF. It's well known that locally the S0 luminosity function is quite
broad, and comparable to E's (Binggeli, Sandage \& Tamman
1988). Further evidence of their contribution to the RSLF can be found
in the scatter in the color magnitude diagram.  Down to $M_B=-16$,
which is well below the magnitude limit of any of the intermediate
redshift cluster surveys, the scatter in the CM relation for the red
population in the nearby Perseus cluster is less than 0.07 mag
(Conselice et al. 2002).  The small amplitude of this scatter is
typically interpreted to mean that there has been a very {\it small
  range} in star-forming histories in this red population.  The lack
of trend in the amplitude with luminosity is also indicative of
galaxies having {\it comparable range} of formation histories. Greater
scatter is seen at fainter magnitudes (Secker \& Harris 1997) where
the dE population begins to dominate, but these depths have not been
probed at intermediate redshift. Similarly, Poggianti et al. (2001)
find that $50-60\%$ of galaxies at all magnitudes have little to no
evidence for any star formation since $z\sim2$.

These lines of evidence indicate one of three possibilities: (1) the
notion that S0's are young is incorrect; (2) the scatter in the
color-magnitude diagram is insensitive to the age-variations under
consideration; or (3) the younger population is well-distributed in
luminosity such that they contribute to scatter (increase it), but not
in a way that drives a significant trend with magnitude. We suspect
the latter scenario is most likely. An obvious next step is to repeat
the RSLF experiment with adequate imaging or kinematic data to
well-distinguish ellipticals from lenticulars in a large sample of
clusters, probing to depths of at least M$^*$+3 and preferably to
M$^*$+5.

\section{Conclusion}

In this paper, we have explored the red sequence luminosity function
in five intermediate redshift galaxy clusters.  The luminosity
functions are measured from deep UBRIz plus narrow band imaging from
the WIYN telescope.  Red sequence galaxies are identified from their
rest frame colors and photometric redshifts within selection radii of
1 Mpc, $0.25 \times R_{200}$, $0.5 \times R_{200}$, and $R_{200}$.
Extensive simulations are performed to assure the quality of the
detection, photometric redshifts, and measurement of the luminosity
function.  The quality of the data is confirmed through the
measurement of the field luminosity function from the off-cluster
sample.

To provide a low redshift comparison sample, we also measured the red
sequence luminosity function in a set of 59 high mass clusters with
data from the SDSS.  The same process for measuring the luminosity
function for the higher redshift cluster was used here with similar
definitions for cluster galaxies and the red sequence.  For both sets
of clusters, we find comparable luminosity functions to those fount in
the literature for previously studied systems.

We have two primary conclusions concerning the RSLF evolution: 

\begin{itemize}
\item $M_B^*$ evolves in a similar manner as the field luminosity
function and has faded by about 0.7 mags over the last 6.5 Gyrs.
However, little evolution is seen between $z=0$ and $z=0.5$ for the
massive clusters.

\item The faint end slope, $\alpha$, shows no indication of evolution
between our low and intermediate redshift samples.  In addition, we
find no relationship between the cluster velocity dispersion,
$\sigma_v$, and $\alpha$ for the high mass clusters.

\end{itemize}

In an extensive comparison to measurements from the literature, we
have two additional conclusions:

\begin{itemize}
\item Selection effects can be critical to the determination of any
signal of evolution.  Clusters do show variations with the luminosity
function in terms of mass and radius, which can lead to erroneous
conclusions in terms of evolution, if not carefully accounted.
Although small, this survey predominately measures the RSLF in massive
system which seems to display different behaviors than low mass
systems.

\item Significant cluster to cluster variations exists, even at a
given mass. The dispersion in cluster luminosity-function parameters,
measured for individual clusters, is typically an order of magnitude
greater than the error estimate on those parameters from fitting to an
ensemble-averaged luminosity function.  Significantly more work needs
to be done to better understand these cluster to cluster variations.

\end{itemize}

A clear picture of the evolution of the cluster RSLF remains elusive,
yet such a picture is critical for understanding the processes that
drive the growth and transformations of cluster-galaxies over cosmic
time. It is conceivable that a sufficiently delineated map of the RSLF
evolution with time, cluster-mass, and location within the local
cluster environment can help confirm or refute predictions of $\Lambda
CDM$ models on time-scales relevant to the assembly of the most
massive, virialized systems in the universe. While the observational
challenge has yet to be met, the theoretical models also fall short
on definitive predictions due to the complexity of the gas-physics
(including star-formation) that channels the transformation of
galaxies from the blue cloud onto the red sequence. For example, clean
predictions of how and when the different sub-populations along the
red sequence (e.g., E, S0, dE) are formed are yet to be had. As
such, the red sequence, as interpreted simply as a mass sequence, will
likely continue as a critical observational foil thrown up as a test
of hierarchical models. Tracking the fate of the blue population of
galaxies that cause the Butchler-Oemler effect is a key to
understanding the physics behind the transformative processes in
clusters; at least some of these systems are likely to be the
progenitors of the red-sequence galaxies. The complexity of the
astrophysics will likely stymie a definitive observational picture of
the transformation process, however headway can be made by juxtaposing
the blue and red populations in the context of environment.  In future
work we will investigate the cluster blue-galaxy population in this
context.

We would like to thank Vy Tran and Chris Moran for providing access to
their spectroscopic redshifts for MS1054 and MS0451, the anonymous
referee for comments that dramatically expanded and improved this
work, SAAO for support (SMC), and U. Toronto for hospitality while
pursuing this work (MAB).  Research was supported by STScI/AR-9917,
NSF/AST-0307417, NSF/AST-0607516 and a Wisconsin Space Grant. We
acknowledge use of the Sloan Digital Sky Survey (SDSS and SDSS-II; see
//www.sdss.org/ for funding, management and participating
institutions).

\appendix

\section{Ensemble-average Luminosity Function \label{app:cum}}

An important aspect of the RSLF brought forward by our work so far is
the variation between clusters seen even at the same redshift and with
similar masses. This phenomenon is fairly well documented for local
cluster samples. A number of studies measure the luminosity function
not for individual clusters but instead for the coadded luminosity
distribution of an ensemble of clusters. Such studies generally either
compute a straight average of the luminosity distribution, or a
weighted average. The primary method for weighting is to normalize
each cluster by the number of bright galaxies (e.g., Colless 1989).

Unfortunately there has been no investigation of whether systematics
in the inferred luminosity-function parameters are introduced by the
cluster-averaging described above, particularly relative to the
measurement of individual clusters. Since there is a natural need to
average over clusters at higher redshift where sources are fainter, it
is important to understand if this averaging will lead to spurious
trends with redshift.

To explore this possibility, we created a number of simulations using
two mock catalogs of 10 and 100 clusters.  We apply errors on the mock
cluster-galaxy counts that are representative of those for our low and
intermediate redshift data.  For each of these two catalogs, we
generate 100 realizations for each luminosity-function distribution we
describe below. For each catalog realization, we measured the
Schechter-function parameters ($\alpha$, M$^*$, and $\Phi^*$) and DGR
(following de Lucia et al. 2007) for each individual cluster, the
average of all the clusters, and the weighted average of the clusters
following the Colless (1989) prescription.

For an initial simulation test-set, we assume all clusters have the
same luminosity function. Unsurprisingly, in this sample, all three
methods return the same results for the luminosity function and DGR.
For the remainder of the simulations, we assume a distribution of
luminosity functions described by Gaussian distribution about
$M^*=-20.0$ and $\alpha=-0.9$. Simulations are generated with
distribution-widths in each of these parameters of 0.05, 0.10, and
0.25 (with appropriate units).  The count-normalization of each
luminosity function, $\Phi^*$ is also assumed to have a Gaussian
distribution with widths corresponding to variations of $20\%$,
$40\%$, and $100\%$ about the mean $\Phi^*$ (with the requirement that
$\Phi^* > 0$). Simulations were carried out where each of the
parameters ($\alpha$, $M^*$, and $\Phi^*$) was varied individually and
also in combination with the other parameters.  When multiple
parameters are varied, the same distribution widths were used for
$\alpha$ and $M^*$, with the corresponding low-, medium-, or
high-percentage width for $\Phi^*$. For example, when $\sigma$ = 0.10
for $\alpha$ and $M^*$, the $\Phi^*$ distribution width is 40\%.  One
further simulation (lowz) uses variations of $0.25$, $0.5$, and $5\%$
in $\alpha$, M$^*$, and $\Phi^*$, respectively, which closes matches
the values found for the low redshift clusters.

The results for the simulations are presented in Table \ref{tab:lfvar}
for realizations including 10 clusters.  The first column lists the
parameter varied ($\alpha$, $M^*$, or $\Phi^*$); the second column
lists the distribution width for that parameter. Subsequent columns
list the average and standard deviation for $\alpha$, $M^*$, and the
DGR measured from the 100 realizations (a) for individual clusters
(columns 3-8), (b) the composite luminosity function constructed by
averaging the cluster-counts together (columns 9-14), and (c) the same
for a composite luminosity function constructed through a weighted
average of the luminosity functions following Colless (1989; columns
15-20 here).  For each quantity, the standard deviation is the
measured deviation about the mean measurement across the 100
simulations.  Simulations including 100 clusters yielded identical
results in the mean, but with smaller standard deviations about the
mean for each value. This is due simply to a better sampling of the
luminosity-function distribution. One exception is for the standard
deviation of the DGR for individually-measured clusters, which are
larger for the 100-cluster simulation (see below).

For measurements of individual clusters, we are able to recover the
original Schechter function values of the parent distribution even for
large spread in the parameterization. In contrast, the measured DGR
does not behave in a similar manner as the parameterization.  For a
single cluster with the nominal luminosity function, the DGR would
have a value of 4.71.  However, as the variation in the luminosity
function parameterization increases, the average value of the DGR
tends to increase as well.  The increase in the average value of the
DGR is due to the Poissian-like distribution in the DGR values due to
a Gaussian spread in $\alpha$.  The median is a better statistic to
measure the DGR from an ensemble of individual measurements than the
average.  For individual measurements of 100 clusters with
$\sigma_{\alpha=0.25}$, the median value of the DGR is 4.78, while the
average is 5.00.

The results for the composite luminosity functions indicate that
averaged luminosity functions tend to yield estimates of increasingly
negative $\alpha$ as the variation in $\alpha$ increases, whereas the
weighted luminosity yields the opposite effect.  In both cases, we see
a small increase in the measured mean value of $M^*$, with the increase
comparable to the measured dispersion.  Neither case shows any change
in parameterization with changes in $\Phi^*$.  For both cases, the DGR
is far better behaved (returning a DGR corresponding to the value
measured for $\alpha$ in each case), although the systematic deviation
in the mean DGR value is no longer coupled with the underlying
variation in luminosity function parameters.  As expected, the
composite luminosity function masks the intrinsic cluster to cluster
variations.

For the lowz case, which represents the dispersion measured for our
low redshift clusters, we find very different results from the
individual, average, and weighted measurements. The individual cluster
measurements do return the input luminosity function as expected, but
with significant variation in $\alpha$ and $M^*$.  The average DGR is
large, but the median, once again, provides a much more accurate
statistic.  However, the average and composite luminosity functions
both perform much worse.  The average luminosity function measures a
far steeper slope of $\alpha=-1.11$ rather than the input value of
$\alpha=-0.90$.  The weighted luminosity function performs slightly
better, but the measured $\alpha$ also decreases to $-0.97$.  Even
worse, the measured $M^*$ increases to $M^*=-20.48$ for both cases
(nearly an 0.5 mag shift). Because of this shift in $M^*$, the
resultant DGR value is generally found to be lower than the nominal
value.

These simulations indicate systematics effects are introduced by
coadding data from an ensemble of clusters. These effects are
minimized by avoiding the DGR formulation and measuring
Schechter-function parameters. In the case where luminosity functions
can be measured for individual clusters, this is clearly
preferable. When this is not the case, we suspect a maximum likelihood
approach that simultaneously fits the cluster ensemble assuming a
distribution in luminosity function-parameters may be promising.

\newpage

\begin{deluxetable}{lrrrrrrr}
\tabletypesize{\tiny}
\tablewidth{0pt}
\tablecaption{Intermediate-Redshift Cluster Properties}
\tablehead{
\colhead{Cluster} & 
\colhead{RA} & 
\colhead{DEC} & 
\colhead{z} & 
\colhead{$\sigma$\tablenotemark{a}} & 
\colhead{$L_x^{bol}$\tablenotemark{b}} & 
\colhead{$R_{200}$} &
\colhead{$M_B^{lim}$} \\
\colhead{} & 
\colhead{(J2000)} &  
\colhead{(J2000)} & 
\colhead{}& 
\colhead{(km s$^{-1}$)} & 
\colhead{erg $s^-1$}& 
\colhead{(Mpc)}& 
\colhead{} \\
}
\startdata
MS0451 & 04:54:10.81 & -03:00:56.85 & 0.54 & 1354 & $40\times10^{44}$  & 2.50 & -17.12 \\
Cl0016 & 00:18:33.52 & +16:26:16.01 & 0.55  & 1230 & $37\times10^{44}$  & 2.25 & -16.70 \\
Cl1322 & 13:24:50.13 & +30:11:18.49 & 0.75  & 1016 & $1.4\times10^{44}$ & 1.66 & -17.59 \\
MS1054 & 10:56:59.50 & -03:37:28.39 & 0.83  & 1170 & $16\times10^{44}$  & 1.82 & -18.45 \\
Cl1604 & 16:04:18.27 & +43:04:38.42 & 0.90  & 982  & $2.0\times10^{44}$ & 1.50 & -18.95 \\
\enddata

\tablenotetext{a}{ Ref: Carlberg et al. 1998, Tran et al. 1999,  Lubin et al. 2004, Gal \& Lubin 2004.}
\tablenotetext{b}{ Ref: Donahue et al. 2003, Henry 2004, Lubin et al. 2002, Gioia et al. 2004}

\label{tab:clust}
\end{deluxetable}

\begin{deluxetable}{lrrrrrrr}
\tabletypesize{\tiny}
\tablewidth{0pt}
\tablecaption{Intermediate-Redshift Cluster Red Sequence Luminosity Functions}
\tablehead{
\colhead{Cluster} & 
\colhead{z} & 
\colhead{R} & 
\colhead{$R_{200}$} &
\colhead{$\alpha$} & 
\colhead{$M_B^*$} & 
\colhead{$\Phi^*$}  &
\colhead{$\chi^2_{\nu}$} \\
\colhead{} &  
\colhead{} & 
\colhead{(Mpc)} & 
\colhead{} & 
\colhead{} & 
\colhead{} & 
\colhead{($10^{-3}$ Mpc$^{-3}$)} & 
\colhead{}\\
}
\startdata
MS0451 & 0.54 & 0.63 & 0.25 & -0.00$^{+0.30}_{-0.24}$  & -20.28$^{+0.30}_{-0.30}$  & 240.0$^{+13.0}_{-36.0}$  & 0.44 \\
MS0451 & 0.54 & 1.00 & 0.40 & -0.22$^{+0.24}_{-0.20}$  & -20.54$^{+0.28}_{-0.28}$  & 130.9$^{+17.1}_{-21.9}$  & 0.69 \\
MS0451 & 0.54 & 1.25 & 0.50 & -0.14$^{+0.24}_{-0.22}$  & -20.44$^{+0.24}_{-0.26}$  & 104.2$^{+10.8}_{-16.2}$  & 1.16 \\
MS0451 & 0.54 & 2.50 & 1.00 & -0.38$^{+0.16}_{-0.14}$  & -20.72$^{+0.20}_{-0.20}$  & 33.6$^{+ 4.4}_{-4.6}$    & 1.22 \\
Cl0016 & 0.55 & 0.56 & 0.25 & -0.38$^{+0.28}_{-0.20}$  & -19.84$^{+0.42}_{-0.44}$  & 250.0$^{+59.0}_{-54.0}$  & 0.76 \\
Cl0016 & 0.55 & 1.00 & 0.44 & -0.58$^{+0.16}_{-0.14}$  & -20.34$^{+0.32}_{-0.34}$  & 110.1$^{+23.9}_{-23.1}$  & 0.98 \\
Cl0016 & 0.55 & 1.12 & 0.50 & -0.60$^{+0.16}_{-0.12}$  & -20.36$^{+0.30}_{-0.30}$  & 99.4$^{+21.6}_{-18.4}$   & 1.14 \\
Cl0016 & 0.55 & 2.25 & 1.00 & -0.82$^{+0.10}_{-0.08}$  & -21.36$^{+0.28}_{-0.26}$  & 25.5$^{+ 5.5}_{-4.5}$    & 1.23 \\
Cl1322 & 0.75 & 0.42 & 0.25 & -0.96$^{+0.64}_{-0.24}$  & -22.80$^{+1.68}_{-0.20}$  & 40.0$^{+56.0}_{-23.0}$   & 0.35 \\
Cl1322 & 0.75 & 0.83 & 0.50 & -0.64$^{+0.48}_{-0.22}$  & -21.78$^{+0.82}_{-0.22}$  & 27.4$^{+16.6}_{-9.4}$    & 0.56 \\
Cl1322 & 0.75 & 1.00 & 0.60 & -0.64$^{+0.32}_{-0.28}$  & -21.16$^{+0.60}_{-0.86}$  & 12.3$^{+37.7}_{ 6.7}$    & 0.37 \\
Cl1322 & 0.75 & 1.66 & 1.00 & -0.64$^{+0.36}_{-0.26}$  & -21.54$^{+0.54}_{-0.66}$  & 14.5$^{+ 6.5}_{-5.5}$    & 1.18 \\
MS1054 & 0.83 & 0.46 & 0.25 & +0.42$^{+0.58}_{-0.84}$  & -20.26$^{+0.42}_{-0.78}$  & 300.0$^{+ 9.0}_{-58.0}$  & 0.76 \\
MS1054 & 0.83 & 0.91 & 0.50 & -0.50$^{+0.40}_{-0.34}$  & -21.20$^{+0.40}_{-0.44}$  & 110.6$^{+31.4}_{-39.6}$  & 1.52 \\
MS1054 & 0.83 & 1.00 & 0.55 & -0.76$^{+0.22}_{-0.20}$  & -21.30$^{+0.30}_{-0.32}$  & 100.0$^{+25.0}_{-31.0}$  & 0.29 \\
MS1054 & 0.83 & 1.82 & 1.00 & -0.58$^{+0.32}_{-0.28}$  & -21.16$^{+0.34}_{-0.36}$  & 35.8$^{+10.2}_{-10.8}$   & 2.29 \\
Cl1604 & 0.90 & 0.37 & 0.25 & -0.86$^{+1.16}_{-0.46}$  & -23.50$^{+2.18}_{-1.00}$  & 30.0$^{+49.0}_{-20.0}$   & 0.01 \\
Cl1604 & 0.90 & 0.75 & 0.50 & -0.20$^{+1.00}_{-0.82}$  & -21.62$^{+0.74}_{-1.70}$  & 46.1$^{+ 4.9}_{-31.1}$   & 0.16 \\
Cl1604 & 0.90 & 1.00 & 0.67 & -0.74$^{+0.54}_{-0.70}$  & -22.28$^{+0.84}_{-1.72}$  & 16.4$^{+-1.4}_{-15.4}$   & 0.54 \\
Cl1604 & 0.90 & 1.50 & 1.00 & -0.94$^{+1.08}_{-0.38}$  & -22.50$^{+1.24}_{-1.50}$  &  6.5$^{+23.5}_{-3.5}$    & 0.40 \\
\enddata
\label{tab:lfclust}
\end{deluxetable}


\begin{deluxetable}{lrrrrrrrrr}
\tabletypesize{\tiny}
\tablewidth{0pt}
\tablecaption{Low Redshift Cluster RSLF within $R_{200}$ and $M_B^*+3$}
\tablehead{
\colhead{Cluster} & 
\colhead{RA} & 
\colhead{DEC} & 
\colhead{z} & 
\colhead{$\sigma$} & 
\colhead{$N_{spec}$} & 
\colhead{$\alpha$} & 
\colhead{$M_B^*$} & 
\colhead{$\Phi^*$}  &
\colhead{$\chi^2_{\nu}$} \\
\colhead{(Abell)} & 
\colhead{(deg)} &  
\colhead{(deg)} & 
\colhead{} & 
\colhead{(km s$^{-1}$)} & 
\colhead{} & 
\colhead{} & 
\colhead{} & 
\colhead{($10^{-3}$ Mpc$^{-3}$)} & 
\colhead{}\\
}
\startdata
0085 & 10.41 & -9.34 & 0.0550 & 1266 & 51 & -0.84$^{+0.10}_{-0.12}$  & -20.52$^{+0.22}_{-0.26}$  & 194.5$^{+37.5}_{-42.5}$  & 1.46 \\
0117 & 14.00 & -10.03 & 0.0548 &  623 & 35 & -0.66$^{+0.30}_{-0.26}$  & -20.06$^{+0.52}_{-0.64}$  & 82.1$^{+31.9}_{-31.1}$  & 1.33 \\
0160 & 18.21 & 15.51 & 0.0434 &  701 & 32 & -0.88$^{+0.42}_{-0.28}$  & -20.38$^{+0.64}_{-1.00}$  & 66.0$^{+10.2}_{-15.8}$  & 0.36 \\
0168 & 18.79 & 0.25 & 0.0450 &  648 & 32 & -0.70$^{+0.22}_{-0.18}$  & -20.10$^{+0.32}_{-0.30}$  & 107.8$^{+30.2}_{-26.8}$  & 0.54 \\
0279 & 29.09 & 1.06 & 0.0800 &  827 & 32 & -1.00$^{+0.20}_{-0.20}$  & -20.54$^{+0.38}_{-0.44}$  & 41.8$^{+18.2}_{-15.8}$  & 1.09 \\
0659 & 126.02 & 19.40 & 0.0987 &  582 & 30 & -0.76$^{+0.16}_{-0.14}$  & -20.70$^{+0.28}_{-0.30}$  & 113.4$^{+28.6}_{-26.4}$  & 0.58 \\
0671 & 127.12 & 30.42 & 0.0503 &  977 & 42 & -1.38$^{+0.22}_{-0.12}$  & -21.50$^{+0.86}_{-0.50}$  & 30.7$^{+40.3}_{-13.7}$  & 1.50 \\
0724 & 134.58 & 38.57 & 0.0933 &  511 & 30 & -1.14$^{+0.12}_{-0.12}$  & -21.54$^{+0.30}_{-0.38}$  & 36.2$^{+12.8}_{-12.2}$  & 0.67 \\
0795 & 141.01 & 14.17 & 0.1390 &  819 & 39 & -0.82$^{+0.12}_{-0.12}$  & -20.90$^{+0.20}_{-0.24}$  & 112.6$^{+22.4}_{-23.6}$  & 1.66 \\
0933 & 151.92 & 0.52 & 0.0970 &  571 & 30 & -0.16$^{+0.42}_{-0.36}$  & -20.06$^{+0.34}_{-0.40}$  & 82.7$^{+ 8.3}_{-20.7}$  & 1.36 \\
0957 & 153.49 & -0.92 & 0.0444 &  766 & 41 & -0.94$^{+0.22}_{-0.20}$  & -20.34$^{+0.42}_{-0.52}$  & 111.8$^{+50.2}_{-43.8}$  & 1.13 \\
1035 & 158.03 & 40.21 & 0.0725 & 1664 & 61 & -0.94$^{+0.10}_{-0.10}$  & -20.84$^{+0.18}_{-0.20}$  & 133.0$^{+27.0}_{-26.0}$  & 1.12 \\
1066 & 159.85 & 5.17 & 0.0688 &  946 & 50 & -0.94$^{+0.14}_{-0.16}$  & -20.68$^{+0.26}_{-0.34}$  & 90.5$^{+25.5}_{-27.5}$  & 1.08 \\
1185 & 167.70 & 28.68 & 0.0324 &  802 & 40 & -1.12$^{+0.26}_{-0.14}$  & -20.36$^{+0.56}_{-0.42}$  & 90.5$^{+65.5}_{-31.5}$  & 1.34 \\
1190 & 167.94 & 40.84 & 0.0754 &  850 & 47 & -0.22$^{+0.20}_{-0.16}$  & -19.82$^{+0.20}_{-0.16}$  & 178.4$^{+15.6}_{-18.4}$  & 0.87 \\
1203 & 168.49 & 40.29 & 0.0752 &  468 & 33 & -0.24$^{+0.44}_{-0.38}$  & -19.76$^{+0.46}_{-0.68}$  & 69.1$^{+ 9.9}_{-21.1}$  & 0.40 \\
1205 & 168.34 & 2.51 & 0.0754 &  743 & 35 & -0.78$^{+0.20}_{-0.18}$  & -20.44$^{+0.32}_{-0.32}$  & 79.8$^{+24.2}_{-21.8}$  & 0.66 \\
1213 & 169.12 & 29.26 & 0.0466 &  664 & 39 & -0.06$^{+0.36}_{-0.34}$  & -19.32$^{+0.30}_{-0.40}$  & 115.6$^{+ 8.4}_{-22.6}$  & 2.42 \\
1291 & 173.02 & 56.02 & 0.0558 & 1272 & 33 & -0.98$^{+0.12}_{-0.12}$  & -20.72$^{+0.26}_{-0.26}$  & 93.1$^{+24.9}_{-23.1}$  & 0.90 \\
1307 & 173.20 & 14.52 & 0.0810 &  927 & 45 & -0.58$^{+0.18}_{-0.18}$  & -20.20$^{+0.24}_{-0.26}$  & 144.8$^{+27.2}_{-30.8}$  & 0.74 \\
1346 & 175.29 & 5.69 & 0.0984 &  732 & 36 & -0.72$^{+0.20}_{-0.18}$  & -20.46$^{+0.28}_{-0.32}$  & 84.9$^{+22.1}_{-22.9}$  & 0.71 \\
1371 & 176.37 & 15.54 & 0.0689 &  611 & 33 & -1.22$^{+0.24}_{-0.18}$  & -21.44$^{+0.74}_{-0.56}$  & 16.5$^{+16.5}_{-8.5}$  & 1.80 \\
1424 & 179.39 & 5.04 & 0.0768 &  573 & 30 & -0.66$^{+0.26}_{-0.22}$  & -20.32$^{+0.34}_{-0.34}$  & 52.6$^{+14.4}_{-14.6}$  & 0.73 \\
1437 & 180.12 & 3.34 & 0.1344 & 1566 & 30 & -1.04$^{+0.10}_{-0.06}$  & -20.78$^{+0.18}_{-0.14}$  & 154.2$^{+32.8}_{-21.2}$  & 1.18 \\
1516 & 184.74 & 5.24 & 0.0766 & 1068 & 30 & -1.04$^{+0.14}_{-0.12}$  & -21.36$^{+0.40}_{-0.46}$  & 59.8$^{+24.2}_{-19.8}$  & 0.75 \\
1541 & 186.86 & 8.84 & 0.0903 &  835 & 45 & -1.00$^{+0.18}_{-0.18}$  & -21.18$^{+0.38}_{-0.50}$  & 57.6$^{+25.4}_{-23.6}$  & 1.41 \\
1552 & 187.46 & 11.74 & 0.0856 &  921 & 47 & -0.84$^{+0.14}_{-0.16}$  & -20.42$^{+0.22}_{-0.26}$  & 124.7$^{+28.3}_{-31.7}$  & 0.87 \\
1650 & 194.69 & -1.75 & 0.0843 &  897 & 31 & -0.74$^{+0.18}_{-0.16}$  & -20.24$^{+0.28}_{-0.34}$  & 150.4$^{+38.6}_{-38.4}$  & 0.74 \\
1663 & 195.69 & -2.52 & 0.0834 &  791 & 41 & -0.78$^{+0.20}_{-0.18}$  & -20.68$^{+0.32}_{-0.38}$  & 76.5$^{+23.5}_{-22.5}$  & 0.61 \\
1691 & 197.85 & 39.20 & 0.0722 &  923 & 34 & -0.58$^{+0.20}_{-0.16}$  & -20.16$^{+0.26}_{-0.26}$  & 127.0$^{+26.0}_{-25.0}$  & 0.59 \\
1750 & 202.72 & -1.84 & 0.0851 & 1062 & 41 & -0.98$^{+0.12}_{-0.12}$  & -20.74$^{+0.26}_{-0.30}$  & 128.5$^{+33.5}_{-32.5}$  & 1.01 \\
1767 & 204.00 & 59.21 & 0.0710 &  888 & 40 & -0.60$^{+0.26}_{-0.22}$  & -20.24$^{+0.36}_{-0.40}$  & 99.5$^{+27.5}_{-28.5}$  & 1.49 \\
1773 & 205.54 & 2.25 & 0.0777 &  998 & 41 & -0.88$^{+0.16}_{-0.18}$  & -20.76$^{+0.28}_{-0.40}$  & 84.3$^{+24.7}_{-27.3}$  & 1.04 \\
1775 & 205.48 & 26.36 & 0.0742 & 1048 & 36 & -1.04$^{+0.22}_{-0.18}$  & -20.66$^{+0.48}_{-0.52}$  & 85.1$^{+47.9}_{-35.1}$  & 1.10 \\
1795 & 207.25 & 26.59 & 0.0630 &  959 & 38 & -0.50$^{+0.24}_{-0.18}$  & -19.70$^{+0.28}_{-0.24}$  & 224.8$^{+43.2}_{-40.8}$  & 1.36 \\
1809 & 208.33 & 5.15 & 0.0792 &  843 & 52 & -1.00$^{+0.18}_{-0.14}$  & -20.84$^{+0.42}_{-0.40}$  & 70.9$^{+32.1}_{-22.9}$  & 1.11 \\
1890 & 214.39 & 8.19 & 0.0578 &  631 & 36 & -1.00$^{+0.24}_{-0.20}$  & -20.70$^{+0.66}_{-0.78}$  & 43.0$^{+29.0}_{-21.0}$  & 1.49 \\
1904 & 215.53 & 48.56 & 0.0712 &  869 & 50 & -0.96$^{+0.18}_{-0.14}$  & -20.88$^{+0.46}_{-0.42}$  & 71.1$^{+31.9}_{-22.1}$  & 1.53 \\
2026 & 227.14 & -0.27 & 0.0906 &  751 & 32 & 0.06$^{+0.52}_{-0.44}$  & -19.26$^{+0.32}_{-0.40}$  & 112.1$^{+ 3.9}_{-21.1}$  & 1.71 \\
2030 & 227.82 & -0.09 & 0.0914 &  534 & 36 & -0.94$^{+0.18}_{-0.14}$  & -21.54$^{+0.60}_{-0.44}$  & 31.1$^{+13.9}_{-10.1}$  & 0.21 \\
2034 & 227.55 & 33.53 & 0.1137 & 1396 & 40 & -1.08$^{+0.10}_{-0.08}$  & -20.98$^{+0.24}_{-0.24}$  & 119.6$^{+32.4}_{-25.6}$  & 0.72 \\
2040 & 228.19 & 7.43 & 0.0460 &  826 & 34 & -1.20$^{+0.18}_{-0.12}$  & -21.10$^{+0.56}_{-0.52}$  & 64.3$^{+44.7}_{-25.3}$  & 1.38 \\
2048 & 228.82 & 4.38 & 0.0980 &  936 & 39 & -0.96$^{+0.06}_{-0.08}$  & -21.16$^{+0.16}_{-0.20}$  & 138.0$^{+20.0}_{-25.0}$  & 0.64 \\
2051 & 229.19 & -0.95 & 0.1195 &  948 & 32 & -0.96$^{+0.10}_{-0.08}$  & -20.54$^{+0.18}_{-0.18}$  & 298.3$^{+59.7}_{-49.3}$  & 1.91 \\
2061 & 230.31 & 30.65 & 0.0773 &  772 & 64 & -0.48$^{+0.22}_{-0.22}$  & -20.02$^{+0.28}_{-0.34}$  & 154.5$^{+29.5}_{-36.5}$  & 1.24 \\
2069 & 230.99 & 29.89 & 0.1137 & 1118 & 50 & -0.70$^{+0.12}_{-0.12}$  & -20.36$^{+0.14}_{-0.16}$  & 169.6$^{+24.4}_{-26.6}$  & 1.49 \\
2094 & 234.15 & -2.03 & 0.1448 &  735 & 30 & -0.94$^{+0.16}_{-0.14}$  & -20.78$^{+0.32}_{-0.34}$  & 80.1$^{+26.9}_{-23.1}$  & 0.62 \\
2108 & 235.02 & 17.89 & 0.0907 &  795 & 32 & -1.00$^{+0.10}_{-0.10}$  & -21.20$^{+0.30}_{-0.30}$  & 74.6$^{+20.4}_{-18.6}$  & 0.75 \\
2122 & 236.12 & 36.13 & 0.0656 &  815 & 33 & -0.54$^{+0.18}_{-0.18}$  & -20.18$^{+0.24}_{-0.26}$  & 107.1$^{+17.9}_{-22.1}$  & 0.89 \\
2124 & 236.25 & 36.06 & 0.0664 & 1260 & 39 & -0.78$^{+0.10}_{-0.12}$  & -20.50$^{+0.20}_{-0.22}$  & 139.2$^{+23.8}_{-27.2}$  & 1.46 \\
2142 & 239.57 & 27.22 & 0.0903 & 1217 & 90 & -0.82$^{+0.06}_{-0.06}$  & -20.56$^{+0.12}_{-0.10}$  & 547.7$^{+59.3}_{-53.7}$  & 1.81 \\
2151 & 241.31 & 17.75 & 0.0344 &  661 & 56 & -1.38$^{+0.04}_{-0.04}$  & -21.32$^{+0.22}_{0.30}$  & 36.2$^{+ 7.8}_{-3.2}$  & 2.61 \\
2175 & 245.10 & 29.92 & 0.0960 &  994 & 34 & -1.06$^{+0.08}_{-0.06}$  & -21.08$^{+0.20}_{-0.18}$  & 145.0$^{+31.0}_{-23.0}$  & 0.62 \\
2244 & 255.68 & 34.05 & 0.1000 & 1138 & 57 & -1.16$^{+0.06}_{-0.06}$  & -20.94$^{+0.14}_{-0.16}$  & 178.1$^{+29.9}_{-29.1}$  & 1.41 \\
2245 & 255.69 & 33.53 & 0.0884 & 1276 & 55 & -1.02$^{+0.06}_{-0.04}$  & -21.14$^{+0.16}_{-0.14}$  & 216.3$^{+33.7}_{-24.3}$  & 1.28 \\
2255 & 258.13 & 64.09 & 0.0803 & 1148 & 45 & -0.50$^{+0.16}_{-0.14}$  & -20.30$^{+0.24}_{-0.20}$  & 186.7$^{+30.3}_{-28.7}$  & 1.41 \\
2399 & 329.39 & -7.79 & 0.0584 &  592 & 43 & -0.56$^{+0.14}_{-0.14}$  & -20.46$^{+0.22}_{-0.20}$  & 102.3$^{+15.7}_{-17.3}$  & 1.13 \\
2593 & 351.13 & 14.64 & 0.0409 &  800 & 36 & -1.18$^{+0.20}_{-0.22}$  & -20.20$^{+0.32}_{-0.40}$  & 49.4$^{+ 16.6}_{-11.4}$  & 1.63 \\
2670 & 358.54 & -10.41 & 0.0757 &  960 & 50 & -0.72$^{+0.14}_{-0.14}$  & -20.58$^{+0.22}_{-0.22}$  & 114.1$^{+22.9}_{-23.1}$  & 1.27 \\
\enddata

\label{tab:lowz}
\end{deluxetable}

\newcommand\tng{\,\tablenotemark{g}}
\newcommand\tnh{\,\tablenotemark{h}}
\newcommand\tni{\,\tablenotemark{i}}
\newcommand\tnj{\,\tablenotemark{j}}
\newcommand\tnk{\,\tablenotemark{k}}

\begin{deluxetable}{lrrrrrrrrrrr}
\tabletypesize{\tiny}
\rotate
\tablewidth{0pt}
\tablecaption{Summary of Extant Cluster RSLF Studies}
\tablehead{
\colhead{Reference} & 
\colhead{z\tablenotemark{a}} & 
\colhead{$N_{clust}$} & 
\colhead{FOV\tablenotemark{b}} & 
\colhead{R\tablenotemark{c}} &
\colhead{$M_{lim}$} & 
\colhead{$M^*-M$} & 
\colhead{Filters} & 
\colhead{Selection} & 
\colhead{Background\tablenotemark{d}} &
\colhead{Combine\tablenotemark{e}} &
\colhead{Measurement\tablenotemark{f}} \\
}
\startdata
Andreon et al. 2006   & 0.0325          & 1  & 1176   & 1.3 Mpc        & 24.5 & 8   & BVR          & B-V       & BGS      & Ind & $M_B$-LF \\
Tanaka et al. 2005    & 0-0.065(0.035)  & --- & ----  & $R_{200}$\tng  & 23 & 3.5 & ugriz\tnh      & $(U-B)_o$ & SZ       & Ave & $M_V$-LF \\
Yagi et al. 2002      & 0-0.076(0.043)  & 10 & 4698   & 1 Mpc          & 20.0 & 6   & R            & $r^{1/4}$ & BGS      & Wei & $M_R$-LF \\
Mercurio et al. 2006  & 0.05   		& 3+ & 14400  & ---- \tng      & 22.0 & 7   & BR           & B-R       & BGS      & Ave & $m_B$-LF \\
de Propris et al. 2003 &0-0.11(0.064)    & 60 & ---    & 1.5 Mpc        & 22.5 &  5  & b$_j$        & pca       & SZ       & Wei & $M_{b_j}-LF$ \\
Low-z clusters (This Study) & 0-0.15 (0.077) & 59  & ----  & $R_{200}$ & 23   & 3,5 & ugriz\tnh    & $(U-B)_o$ & PZ+BGS   & Ind & $M_B$-LF \\
Barkhouse et al. 2007 & 0-0.18(0.08)    & 57 &  529   & $0.4R_{200}$   & 23   & 6.0 & BRI          & B-R       & BGS      &Ind/Ave & $M_R$-LF \\
Stott et al. 2007     & 0.08-0.15(0.12) & 10 & ----   & 0.6 Mpc        & 20.9 & 3.1 & BR           & U-V       & BGS      & Ave & $M_V$-LF/DGR \\
Popesso et al. 2006   & 0-0.460(0.13)   & 59 & ----   & $R_{200}$      & 23   & 6.5 & ugriz\tnh    & u-r       & BGS      &Ind/Wei& g-LF \\
Goto et al.  2002      & 0.02-0.25(0.18) &204 & ---    & 0.75 Mpc       & 23   &  3.5 & ugriz\tnh   & Morph     & PZ+BGS      & Wei & $M_g-LF$\\
Smail et al. 1998     & 0.22-0.28(0.24) & 10 &  90    &  0.7 Mpc       & 22.8 & 4   & UBI          & B-I       & BGS      & Ave & I-LF \\ 
Muzzin et al. 2006    & 0.35-0.47(0.296)& 15 & 45-250 & $R_{200}$      & 24   & $K^*+2.5$ & grK    & pca       & SZ       & Ave & K-LF \\ 
Wilman et al. 2005    & 0.35-0.47(0.43) & 26 &  30.5+ & 0.5 Mpc        & 22.0 & $\sim 2$  & UBVRI  & EW        & SZ       & Ave & $M_B$ DGR    \\
Gilbank et al. 2007   & 0.40            & 57 & ----   & $0.5R_{200}$   & 23.2 & 4    & Rz          & R-z       & BGS      & Wei & $M_V$-LF  \\     
Barrientos and Lilly 2003 & 0.39-0.48(0.45) & 8 & 213.2 & 1 Mpc        & 24.5 & 3.5 & VIK          & V-I       & BGS      & All & I-LF\\
Stott et al. 2007     & 0.47-0.68(0.54) & 10 & 11.3   & 0.6 Mpc        & 25.3 & 3   & Vi\tni       & U-V       & BGS      & Ave & $M_V$-LF/DGR \\
MS0451 (This study)   & 0.54            &  1 & 92.16  & $R_{200}$      & 25.21 & 3.5 & UBRIz+NB    & $(U-B)_o$ & PZ+BGS   & Ind & $M_B$-LF \\
Cl0016 (This study)   & 0.55            &  1 & 92.16  & $R_{200}$      & 25.62 & 4.6 & UBRIz+NB    & $(U-B)_o$ & PZ+BGS   & Ind & $M_B$-LF \\
de Lucia  et al. 2006 & 0.40-0.80(0.59) & 18 & 42.45  & $0.5R_{200}$   & 25.7 & 2.5 & VRI          & V-I       & PZ+BGS   & Ind/Ave& $M_V$-DGR  \\
Tanaka et al. 2005    & 0.55,0.83(0.69) &  2 & 768    & $R_{200}$\tng  & 26.0 & 3   & BVRiz        & $(U-B)_o$ & PZ+BGS   & Ind & $M_V$-LF \\
Gilbank et al. 2007   & 0.5-0.95        & 101& ----   &  $0.5R_{200}$  & 23.2 & 1-3    & Rz        & R-z       & BGS      & Wei & $M_V$-DGR  \\ 
Cl1322 (This study)   & 0.75            &  1 & 92.16  & $R_{200}$      & 25.56 & 3.9 & UBRIz+NB    & $(U-B)_o$ & PZ+BGS   & Ind & $M_B$-LF \\
Andreon 2008          & 0.50-1.27(0.76) & 14 & 11.33  & ~0.5 Mpc       & 25.5  & 3.5+ & Vi\tni     & $(U-V)_o$ & BGS      & Ind & $M_V$-LF \\
Koyama et al. 2007    & 0.81            &  1 & 768    & $0.35R_{200}$  & 26.0 & 2.0 & VRiz         & R-z       & BGS      & Ind & DGR \\  
Goto et al. 2005      & 0.83            &  1 & 36.51  & 1 Mpc          & 24.5 & 2.5 & Viz\tni      & i-z       & SZ       & Ind & $M_i$-LF  \\
Andreon 2006          & 0.83            &  1 & 20.25  & ~1 Mpc         & 26.5 & 3.5 & ViK\tni      & V-I       & BGS      & Ind & $M_B$-LF \\
MS1054 (This study)   & 0.83            &  1 & 92.16  & $R_{200}$      & 25.48 & 2.7 & UBRIz+NB    & $(U-B)_o$ & PZ+BGS   & Ind & $M_B$-LF \\
Cl1604 (This study)   & 0.90            &  1 & 92.16  & $R_{200}$      & 25.42 & 3.5 & UBRIz+NB    & $(U-B)_o$ & PZ+BGS   & Ind & $M_B$-LF \\
Andreon et al. 2008   & 1.02            &  1 & 49.0   & 0.7 Mpc        & 24.5 & 2    &  Iz         & I-z       & BGS      & Ind  &$M_V$-LF \\
Mei et al. 2006\tnj   & 1.106           &  1 & 11.9   & 0.6 Mpc        & 25.5 & $>1$ &  izJK\tni   & i-z       & PZ       & Ind & $M_B$-LF \\
Stazzullo et al. 2006 & 1.11-1.27(1.17) &  1 & 1.8-4.4 & 0.7-1.0 Mpc   & ---  & $K^*+1.5$ & K      & TEM       & BSG      & Ind & K-LF \\ 
Nakata et al. 2001    & 1.2             &  1 & 0.66   & 0.33 Mpc       & 26   & $K^*+3.0$ & VRIK   & ---       & PZ       & Ind & none \\
Toft et al. 2004      & 1.237           &  1 & 16     & 1 Mpc          & 26   & $K^*+3.5$ & BVRIJK & ---       & PZ       & Ind & K-LF \\
Blakeslee et al. 2003 & 1.24            &  1 & 36.51  & 1 Mpc          & 26.5 & 1.3 & iz\tni       & i-z       & Morph    & Ind & none \\
Tanaka et al. 2007    & 1.24            &  1 & 25     & 0.5-1 Mpc      & 26.4 & $K^*+1.5$ & VrizK  & i-K       & none     & Ind & K-LF \\
Mei et al. 2006\tnk   & 1.265           &  1 &  37    & 0.5 Mpc        & 26.5 & $>2$ &  iz\tni     & i-z       & SZ+morph & Ind & $M_B$-LF \\ 
\enddata

\clearpage

\label{tab:survey}
\tablenotetext{a}{For studies with a single cluster, gives the redshift of the cluster.  For studies with 
multiple clusters, gives the range and median value of the redshifts.}
\tablenotetext{b}{The Field of View (FOV) is in square arcminutes}
\tablenotetext{c}{Limiting magnitude in the R band.  Transformations are made to the R-band from other 
optical bands if necessary.  Other magnitudes indicated when necessary.  All magnitudes transformed to Vega. }
\tablenotetext{d}{Background refers to the method used for the seperation of field and cluster sources.  
The diferent techniques are:  'BGS' is background subtraction, 'SZ' is spectroscopic identification, 
'PZ' is photometric redshift identification, and 'Morph' is using morphological information to identify 
cluster objects.}
\tablenotetext{e}{Combine refers to how the clusters were combined.  The different methods are: 'Ind' means 
the luminosity function was measured for each individual cluster, 'Ave' means a single luminosity function fit to the 
average or sum of all of the clusters, and 'Wei' means a single luminosity function to a weighted average of the clusters.}
\tablenotetext{f}{Type of measurement made to assess the density of objects.  'LF' is luminosity function. 
'DGR' is for dwarf to galaxy ratio, but the definition differs between the different studies.  Before each 
definition, the band (rest or observed) that the measurement is made in is labelled.}
\tablenotetext{g}{Tanaka et al. use a density estimate to define clusters vs. field but
the estimate corresponds to the viral radius, which is approximately equal to $R_{200}$.  Mercurio et al. also use
a density estimate to differentiate between low and high-density regions.}
\tablenotetext{h}{SDSS magnitudes.}
\tablenotetext{i}{HST studies where V, i, and z are the F606W, F775W, and F850LP bands, respectively.  For Stott et al. 2007, the HST
bands for V and i are F555W and F814W. For Andreon 2008, there is only two HST bands that bracket the 4000 \AA break.  Filters used
include F555W, F606W, F775W, F814W, and F850LP.  }
\tablenotetext{j}{Mei et al. 2006, ApJ 639, 81}
\tablenotetext{k}{Mei et al. 2006, ApJ 644, 759.  Observations of a supercluster located at z=1.265.  The two clusters
are separated by 2 Mpc and have redshifts of z=1.26,1.27.}

\end{deluxetable}

\newpage


\begin{deluxetable}{lr|rrrrrr|rrrrrrrr|rrrrrr}
\tabletypesize{\tiny}
\rotate
\tablewidth{0pt}
\tablecolumns{22}
\tablecaption{Simulations of Luminosity-Function Measurement Variations for a 10-Cluster Ensemble}
\tablehead{
\colhead{Parameter\tablenotemark{a}} &
\colhead{$\sigma$\tablenotemark{b}} &
\multicolumn{6}{c}{Individual}  & \colhead{} &
\multicolumn{6}{c}{Average} & \colhead{} &
\multicolumn{6}{c}{Weighted} \\  \cline{3-8} \cline{10-15} \cline{17-22}
\colhead{} & \colhead{} &
\colhead{$\alpha$} &
\colhead{$\sigma_{\alpha}$} &
\colhead{$M^*$} &
\colhead{$\sigma_{M^*}$} &
\colhead{$DGR$} &
\colhead{$\sigma_{DGR}$}  & \colhead{} &
\colhead{$\alpha$} &
\colhead{$\sigma_{\alpha}$} &
\colhead{$M^*$} &
\colhead{$\sigma_{M^*}$} &
\colhead{$DGR$} &
\colhead{$\sigma_{DGR}$} & \colhead{} &
\colhead{$\alpha$} &
\colhead{$\sigma_{\alpha}$ }&
\colhead{$M^*$} &
\colhead{$\sigma_{M^*}$} &
\colhead{$DGR$} &
\colhead{$\sigma_{DGR}$} \\}
\startdata
$\alpha$ & 0.05 & -0.90 & 0.02 & -20.00 & 0.00 & 4.73 & 0.30 &   & -0.91 & 0.02 & -20.01 & 0.01 & 4.72 & 0.11  &   & -0.90 & 0.02 & -20.01 & 0.01 & 4.69 & 0.11 \\
$\alpha$ & 0.10 & -0.90 & 0.03 & -20.00 & 0.00 & 4.80 & 0.67 &   & -0.92 & 0.03 & -20.01 & 0.01 & 4.77 & 0.21  &   & -0.89 & 0.03 & -20.01 & 0.01 & 4.59 & 0.20 \\
$\alpha$ & 0.25 & -0.89 & 0.07 & -20.00 & 0.00 & 4.95 & 1.71 &   & -0.97 & 0.08 & -20.07 & 0.04 & 4.77 & 0.52  &   & -0.82 & 0.07 & -20.06 & 0.03 & 3.93 & 0.48 \\
$\alpha$, $M^*$ & 0.05 & -0.90 & 0.02 & -20.00 & 0.01 & 4.74 & 0.42  &   & -0.91 & 0.02 & -20.01 & 0.02 & 4.72 & 0.14  &   & -0.90 & 0.02 & -20.01 & 0.02 & 4.67 & 0.14 \\
$\alpha$, $M^*$ & 0.10 & -0.90 & 0.03 & -20.00 & 0.03 & 4.82 & 0.84  &   & -0.92 & 0.03 & -20.03 & 0.03 & 4.72 & 0.27  &   & -0.90 & 0.03 & -20.03 & 0.03 & 4.55 & 0.27 \\
$\alpha$, $M^*$ & 0.25 & -0.89 & 0.08 & -20.01 & 0.08 & 5.31 & 2.42  &   & -1.02 & 0.09 & -20.18 & 0.11 & 4.65 & 0.72  &   & -0.86 & 0.09 & -20.16 & 0.14 & 3.79 & 0.76 \\
$\alpha$,$M^*$,$\Phi^*$ & 0.05 & -0.90 & 0.01 & -20.00 & 0.02 & 4.73 & 0.40  &   & -0.90 & 0.02 & -20.01 & 0.02 & 4.70 & 0.15  &   & -0.90 & 0.01 & -20.00 & 0.02 & 4.66 & 0.14 \\
$\alpha$,$M^*$,$\Phi^*$ & 0.10 & -0.90 & 0.03 & -20.00 & 0.04 & 4.79 & 0.85  &   & -0.92 & 0.04 & -20.03 & 0.04 & 4.69 & 0.35  &   & -0.89 & 0.03 & -20.03 & 0.04 & 4.53 & 0.34 \\
$\alpha$,$M^*$,$\Phi^*$ & 0.25 & -0.89 & 0.08 & -20.01 & 0.08 & 5.30 & 2.40  &   & -1.01 & 0.11 & -20.18 & 0.13 & 4.67 & 0.81  &   & -0.87 & 0.08 & -20.18 & 0.14 & 3.77 & 0.70 \\
$\alpha$,$\Phi^*$ & 0.05 & -0.90 & 0.02 & -20.00 & 0.00 & 4.73 & 0.33 &   & -0.91 & 0.02 & -20.01 & 0.01 & 4.72 & 0.10  &   & -0.90 & 0.02 & -20.01 & 0.01 & 4.68 & 0.11 \\
$\alpha$,$\Phi^*$ & 0.10 & -0.90 & 0.03 & -20.00 & 0.00 & 4.73 & 0.66  &   & -0.91 & 0.04 & -20.01 & 0.01 & 4.71 & 0.25  &   & -0.88 & 0.03 & -20.01 & 0.01 & 4.54 & 0.22 \\
$\alpha$,$\Phi^*$ & 0.25 & -0.90 & 0.07 & -20.00 & 0.00 & 5.07 & 1.73  &   & -0.99 & 0.11 & -20.08 & 0.05 & 4.94 & 0.71  &   & -0.84 & 0.08 & -20.06 & 0.03 & 3.98 & 0.54 \\
$M^*$ & 0.05 & -0.90 & 0.00 & -20.00 & 0.02 & 4.74 & 0.26  &   & -0.90 & 0.00 & -20.00 & 0.02 & 4.72 & 0.09  &   & -0.90 & 0.00 & -20.00 & 0.02 & 4.72 & 0.09 \\
$M^*$ & 0.10 & -0.90 & 0.00 & -20.00 & 0.03 & 4.78 & 0.54  &   & -0.91 & 0.01 & -20.01 & 0.03 & 4.71 & 0.18  &   & -0.91 & 0.01 & -20.01 & 0.03 & 4.71 & 0.18 \\
$M^*$ & 0.25 & -0.90 & 0.00 & -19.98 & 0.08 & 5.15 & 1.55 &   & -0.94 & 0.02 & -20.09 & 0.09 & 4.63 & 0.42  &   & -0.94 & 0.02 & -20.08 & 0.09 & 4.64 & 0.42 \\
$M^*$,$\Phi^*$ & 0.05 & -0.90 & 0.00 & -20.00 & 0.02 & 4.72 & 0.25  &   & -0.90 & 0.00 & -20.00 & 0.02 & 4.70 & 0.08  &   & -0.90 & 0.00 & -20.00 & 0.02 & 4.70 & 0.08 \\
$M^*$,$\Phi^*$ & 0.10 & -0.90 & 0.00 & -20.00 & 0.03 & 4.78 & 0.51  &   & -0.90 & 0.01 & -20.00 & 0.03 & 4.72 & 0.17  &   & -0.90 & 0.01 & -20.01 & 0.03 & 4.71 & 0.16 \\
$M^*$,$\Phi^*$ & 0.25 & -0.90 & 0.00 & -19.99 & 0.08 & 5.11 & 1.47  &   & -0.94 & 0.02 & -20.08 & 0.10 & 4.65 & 0.52  &   & -0.94 & 0.02 & -20.09 & 0.09 & 4.61 & 0.43 \\
$\Phi^*$ & 20\% & -0.90 & 0.00 & -20.00 & 0.00 & 4.71 & 0.00  &   & -0.90 & 0.00 & -20.00 & 0.00 & 4.71 & 0.00  &   & -0.90 & 0.00 & -20.00 & 0.00 & 4.71 & 0.00 \\
$\Phi^*$ & 40\% & -0.90 & 0.00 & -20.00 & 0.00 & 4.71 & 0.00  &   & -0.90 & 0.00 & -20.00 & 0.00 & 4.71 & 0.00  &   & -0.90 & 0.00 & -20.00 & 0.00 & 4.71 & 0.00 \\
$\Phi^*$ & 100\% & -0.90 & 0.00 & -20.00 & 0.00 & 4.71 & 0.00  &   & -0.90 & 0.00 & -20.00 & 0.00 & 4.71 & 0.00  &   & -0.90 & 0.00 & -20.00 & 0.00 & 4.71 & 0.00 \\
lowz\tablenotemark{c} & $\cdots$ & -0.90 & 0.07 & -20.03 & 0.14 & 7.10 & 6.60  &   & -1.11 & 0.10 & -20.48 & 0.24 & 4.16 & 0.77  &   & -0.97 & 0.10 & -20.48 & 0.27 & 3.37 & 0.79 \\
\enddata
\tablenotetext{a}{Parameter refers to the luminosity function parameter which is being varied in each simulation.}
\tablenotetext{b}{$\sigma$ refers to the amount the parameter is being varied in the simulation.}
\tablenotetext{c}{``lowz'' is a simulation with variations matching our low-redshift cluster sample.  The simulation has variations in $\alpha$, $M^*$, and $\Phi^*$ of $\sigma_{\alpha}=0.25$, $\sigma{M^*}=0.5$, and $\sigma_{\Phi^*}=5\%$.}
\label{tab:lfvar}
\end{deluxetable}

\clearpage

\onecolumn

\begin{figure}[hbt]
\epsscale{0.90}
\plotone{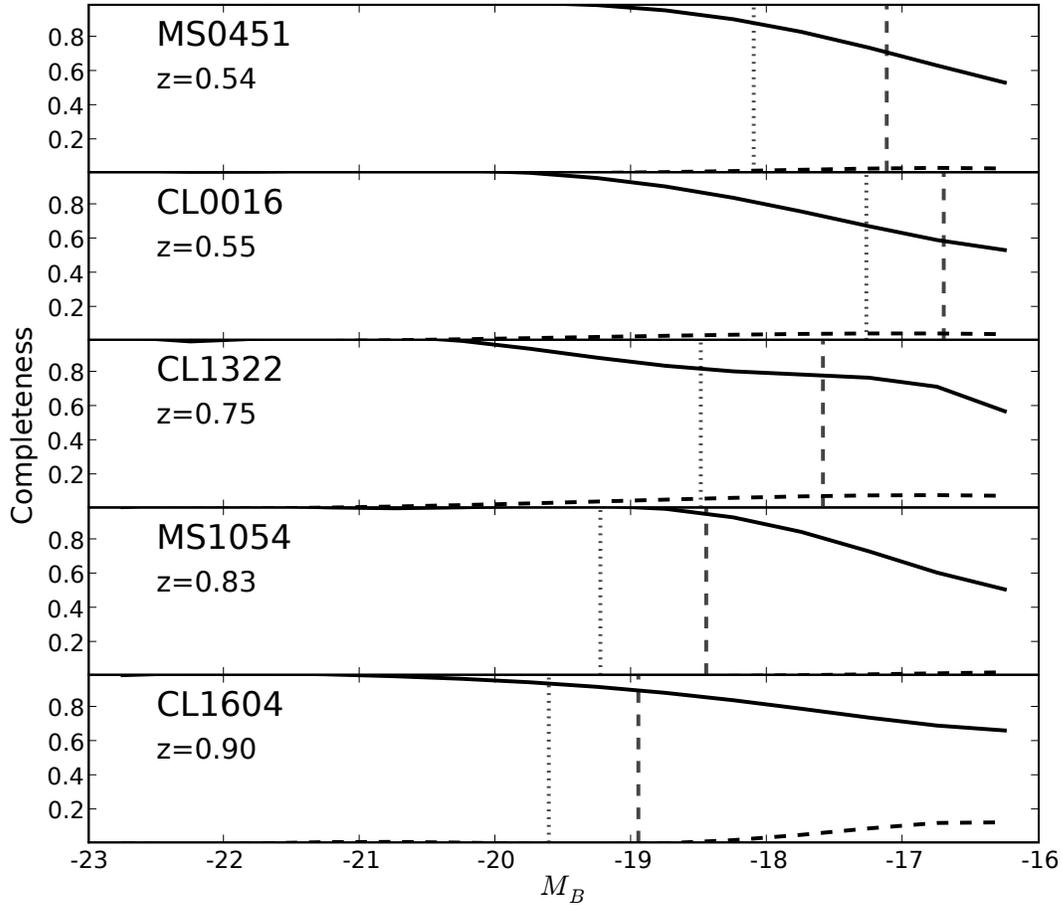}
\caption{Completeness and reliability of red cluster-galaxy selection
  due to effects from photometric redshift errors.  Solid lines
  represent the fraction of simulated cluster galaxies that have
  photometric redshifts within the selection window for the
  cluster. Simulations are described in text. In the worse case,
  corrections for incompleteness due to photometric-redshift errors
  are 40\% at the 50\% detection-completeness limit, and typically are
  only 20-25\%. At the 90\% detection-completeness limit, the
  correction for selection incompleteness is typically below 15\%.
  Dotted-lines represent the fraction of galaxies in the cluster
  luminosity function that are expected to be contamination from the
  field.  Due to the much larger red galaxy luminosity function
  normalization, only a small fraction ($<$5\%) of field galaxies are
  expected to contribute even at the faintest magnitude bins.}
\label{fig:selcomp}
\end{figure}

\begin{figure}[hbt]
\epsscale{0.90}
\plotone{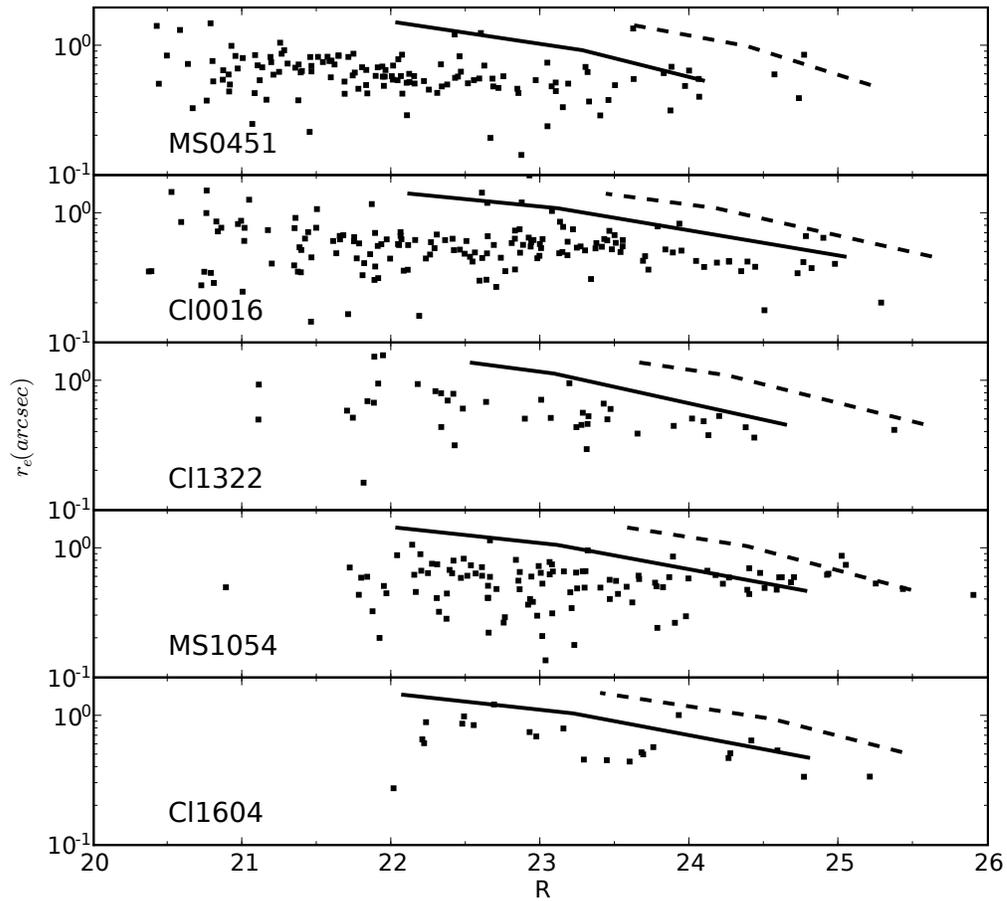}
\caption{Apparent size-magnitude locus for the red cluster objects is
  well offset from the $50\%$ (dashed line) and $90\%$ (solid line)
  completeness limit except for the highest redshift clusters (MS1054,
  and Cl1604).  In these clusters, we may be missing a portion of the
  dimmest objects which would bias our luminosity functions towards a
  deficit.  However, since our completeness corrections include
  corrections for object size and magnitude, this bias is diminished.}
\label{fig:sbias}
\end{figure}

\begin{figure}[hbt]
\epsscale{0.9}
\plotone{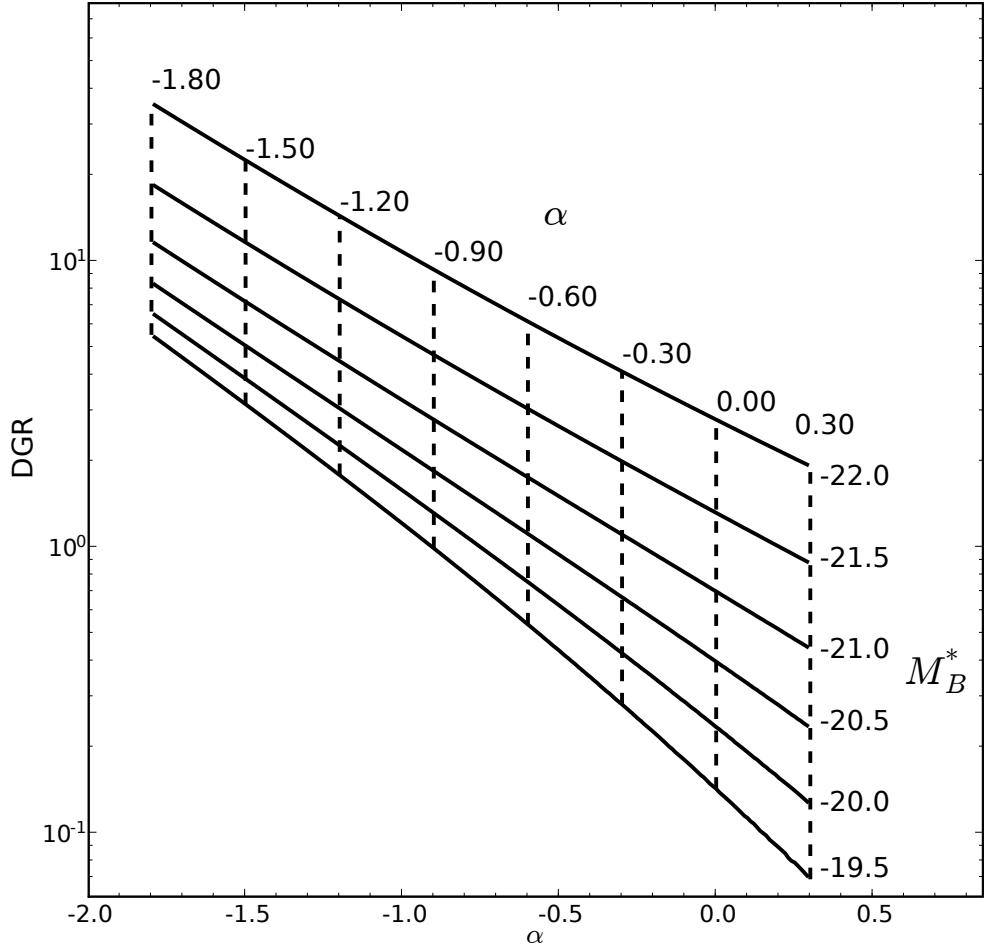}
\caption{Dwarf-to-giant ratio (DGR) relationship with Schechter
  luminosity-function parameters $\alpha$ (labels at top and x axis)
  and $M^*$ (labeled curves at right).  The DGR is primarily a
  function of $\alpha$, but is also affected by the value of $M^*$,
  and hence DGR measurement interpretation is degenerate between
  variations in both these quantities.}
\label{fig:dgralpha}
\end{figure}

\begin{figure}[hbt]
\epsscale{0.8}
\plotone{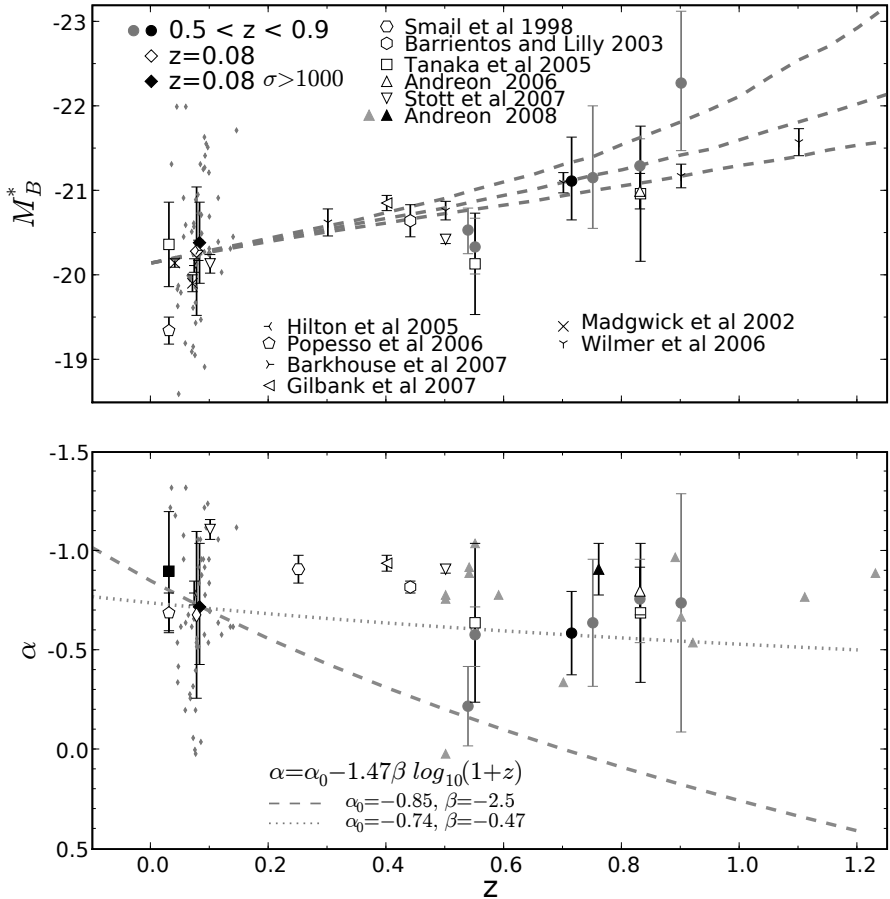}
\caption{Evolution of RSLF $M^*$ (top) and slope (bottom) with
  redshift. Individual data from our survey is marked by filled, gray
  circles and diamonds, whereas the average values are marked by dark
  circles and diamonds (see key).  Other studies are plotted as given
  in the key. All except two field studies (Madgwick et al. 2002,
  Willmer et al. 2006) are for cluster samples.  Dashed curves in top
  panel show evolution in $M_B$ of a solar-metalicity, Salpter-IMF
  simple stellar population with formation epochs of $z = 1.5,2,3$
  (top to bottom), referenced to Madgwick et al. at $z=0$. The dashed
  curve in the bottom panel represents the observed evolution in the
  DGR relationship found by Stott et al. (2007) normalized to our
  low-redshift measurement of $\alpha$ (see text). $M^*$ rises much
  more smoothly with redshift in the field (Madgwick et al. 2002,
  Willmer et al. 2006) than in the cluster sample, whereas $\alpha$
  shows little to trend with redshift.}
\label{fig:zev}
\end{figure}

\begin{figure}[hbt]
\epsscale{0.9}
\plotone{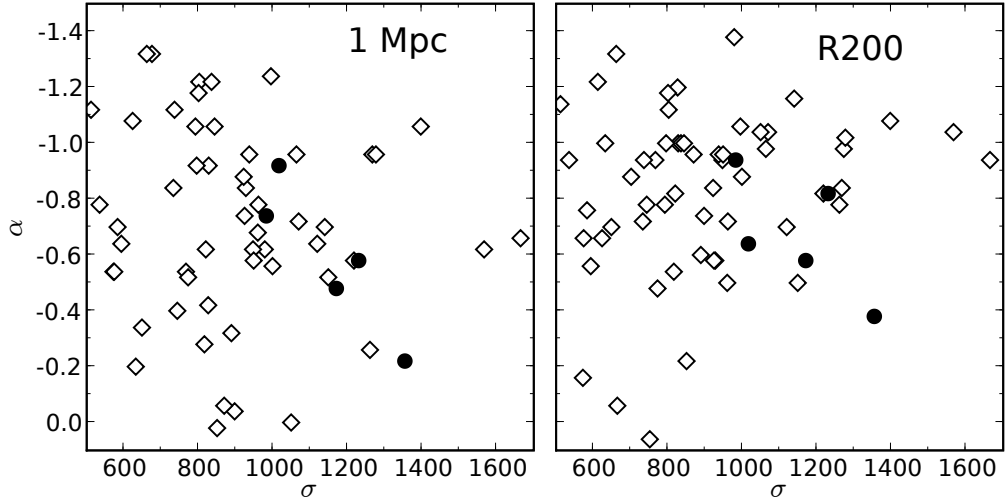}
\caption{Faint end slope, $\alpha$, as a function of cluster velocity
dispersion for low and intermediate redshift clusters.  The
low-redshift data is from measurements based on SDSS spectroscopic
measurements of the Abell sample for magnitudes limits of $M^*+3$.  In
the left panel all data are measured within a selection radius of 1
Mpc. The right panel shows the same, measured within a selection
radius of R$_{200}$.}
\label{fig:massalpha}
\end{figure}

\begin{figure}[hbt]
\epsscale{0.8}
\plotone{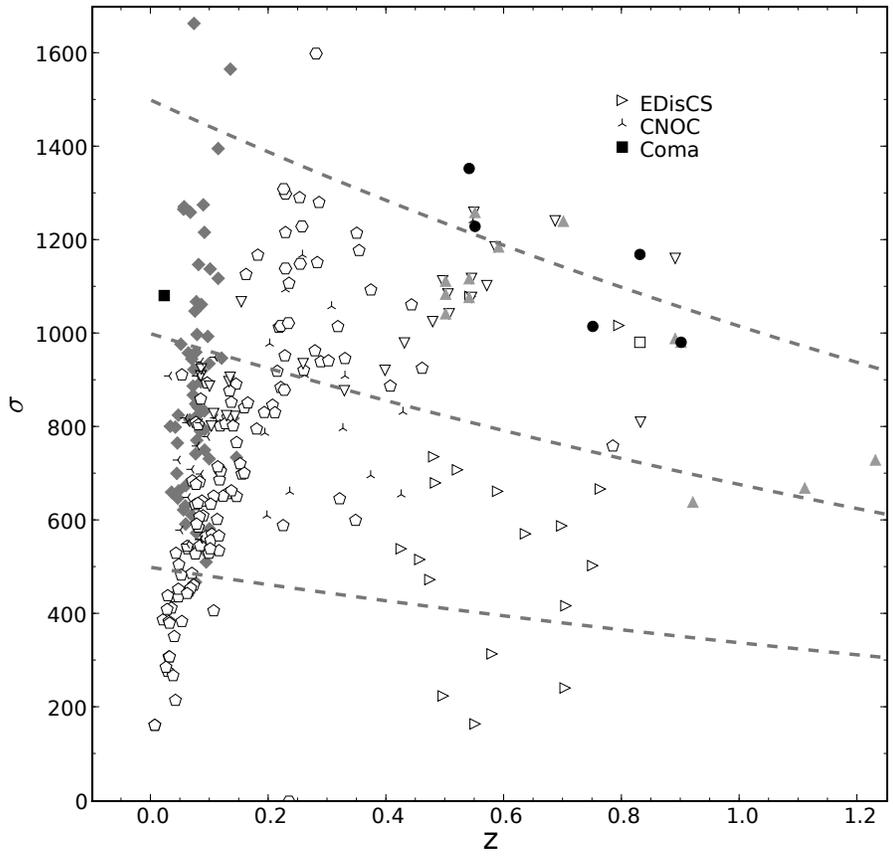}
\caption{Cluster velocity dispersions (masses for constant size) as a
function of redshift.  Our smaller sample of clusters are far more
massive than either the CNOC sample (Yee et al. 1996, Carlberg et
al. 1997; open shields), the EDisCS samples (White et al. 2005, de
Lucia et al. 2006; open diamonds), or even Coma (Colless \& Dunn 1996;
filled square).  Other points are the same as in Figure
\ref{fig:zev}. Predictions for the growth of halos from Wechsler et
al. (2002) are over plotted as dashed line for halos with
$\sigma=1500,1000,500$ km s$^{-1}$ at $z=0$. }
\label{fig:zmass}
\end{figure}

\end{document}